\documentclass[reprint,
 amsmath,amssymb,
 aps,
prb,
]{revtex4-1}

\usepackage{graphicx}
\usepackage{dcolumn}
\usepackage{bm}
\usepackage{longtable}
\usepackage{tabularx}
\usepackage{multirow}
\usepackage{array}
\usepackage{hyperref}

\begin{document}

\preprint{APS/123-QED}

\title{Physical properties of \textit{R}Ir$_3$ (\textit{R} = Gd, Tb, Ho) compounds with coexisting polymorphic phases}

\author{Binita Mondal$^{1, 2}$, Shovan Dan$^{3}$, Sudipta Mondal$^{1, 4}$, R. N. Bhowmik$^{5}$, R. Ranganathan$^{1}$, Chandan Mazumdar$^{1}$}
\email{$^{*}$chandan.mazumdar@saha.ac.in}
\affiliation{$^{1}$Condensed Matter Physics Division, Saha Institute of Nuclear Physics, 1/AF Bidhannagar, Kolkata 700064, India}
\affiliation{$^{2}$K.K.M. College, Jamui, Bihar 811307, India}
\affiliation{$^{3}$Department of Physics, The University of Burdwan, Burdwan 713104, West Bengal, India}
\affiliation{$^{4}$K.S.S College, Lakhisarai, Bihar 811311, India}
\affiliation{$^{5}$Department of Physics, Pondicherry University, R.V. Nagar, Kalapat, Pondicherry 605014, India}

\date{\today}

\begin{abstract}
The binary compounds GdIr$_3$, TbIr$_3$ and HoIr$_3$ are synthesized successfully and found to form in macroscopic co-existence of two polymorphic phases: C15b and AuCu$_3$-type. The dc magnetization and heat capacity studies confirm that C15b phase orders ferromagnetically, whereas the AuCu$_3$ phase remains paramagnetic down to 2 K. The frequency dependent ac-susceptibility data, time dependent magnetic relaxation behavior and magnetic memory effect studies suggest that TbIr$_3$ and HoIr$_3$ are cannonical spin-glass system, but no glassy feature could be found in GdIr$_3$. The critical behavior of all the three compounds has been investigated from the magnetization and heat capacity measurements around the transition temperature (\textit{T}$\rm_{C}$). The critical exponents $\alpha$, $\beta$, $\gamma$ and $\delta$ have been estimated using different techniques such as Arrott-Noaks plot, Kouvel-Fisher plot, critical isotherm as well as analysis of specific heat data and study of magnetocaloric effect. The critical analysis study identifies the type of universal magnetic class in which the three compounds belong.

\end{abstract}

\pacs{Valid PACS appear here}
\maketitle


\section{Introduction}

Physical properties of a material depend strongly on its states, \textit{viz.} solid, liquid, gas, \textit{etc.} It is the density of the constituent particles, or rather the relative arrangement of atoms that determines the state of the matter. One may find different arrangement of atoms in the same state of matter, \textit{e.g.}, in solid state itself, which is arguably the most common form for a large number of materials in normal temperature and pressure. One such example is solid elemental carbon that can exist with different atomic arrangements, \textit{viz.}, graphite, diamond, fullerene (C$_{60}$), C$_{70}$, nanotubes, \textit{etc.} \cite{1, 2, 2a, 2b, 2c, 2d, 2e, 2f} and all these allotropic forms of carbon are known to exhibit a wide range of diverse physical properties \cite{1, 2, 2a, 2b, 2c, 2d, 2e, 2f}. It is not only the elemental solids, but many multi-element compounds are also known to exhibit different properties depending on their relative atomic arrangements. For example, while LaIr$_{2}$Si$_{2}$ forming in CaBe$_{2}$Ge$_{2}$-type crystal structure exhibit superconductivity, the same compound forming in ThCr$_{2}$Si$_{2}$-type crystal structure shows no traces of it down to lowest measureable temperature \cite{3}. On the other hand, PrIr$_{2}$Si$_{2}$, forming in ThCr$_{2}$Si$_{2}$-type structure, exhibit antiferromagnetic ordering at low temperature, but the compound remain paramagnetic when forms in CaBe$_{2}$Ge$_{2}$-type structure \cite{4}. The different atomic arrangements in the respective crystal structures have been argued to be the reason behind their different properties. Such a characteristic, where the chemical composition remains conserved yet having different crystal structure, is commonly known as polymorphism. Except a very few cases like \textit{R}Ir$_{2}$Si$_{2}$ (\textit{R} = rare earth) mentioned above, where the structural change is achieved by annealing the material at high temperature, most often one realize polymorphic phases by tuning external parameters, \textit{viz.}, temperature, pressure, \textit{etc}. However, an external parameter driven structural change does not allow us to compare physical properties under identical conditions, \textit{e.g.}, under the influence of same temperature, pressure, magnetic and electric field, \textit{etc.} Therefore, to study the effect of polymorphism on physical properties, it is highly desirable to obtain different phases under similar environment, \textit{viz.}, temperature, pressure, \textit{etc.} Similar to \textit{R}Ir$_{2}$Si$_{2}$ series of compounds, it has been recently shown that \textit{R}Pt$_{3}$B-type (\textit{R} = rare earth) of compounds that form in tetragonal crystal structure, changes to cubic perovskite like structure when annealed at high temperature \cite{5}, although the structural change found to be associated with creation of partial vacancy in the body-center position. One may note here that the different members of binary \textit{R}Pt$_{3}$ (\textit{R} = rare earth) compounds are known to form in two different crystal structures: C15b type for \textit{R} = La-Tb, and AuCu$_{3}$-type for \textit{R} = Dy-Lu \cite{8, 9, 12, 13}. Additionally, it was found that C15b type of structure in TbPt$_{3}$ appear to be metastable in nature, as it changes to AuCu$_{3}$ type on annealing \cite{8}. Another binary series, \textit{R}Rh$_{3}$ (\textit{R} = rare earth), are also reported to form in different crystal structures (CeNi$_{3}$-, PuNi$_{3}$-, AuCu$_{3}$-types) for different rare earth analogues. Only LaRh$_{3}$ have been reported to form in two different structures \cite{10, 11, 6}. In such scenario, one may expect multiple crystal structures in \textit{R}Ir$_{3}$ series too, since Ir has very similar outer electron configuration as that in Rh, and has only one electron less than that  in Pt. In our work we found that although many members of \textit{R}Ir$_3$ series of compounds could be synthesized with single chemical compositions, but with two different crystal structures coexisting together. We also report here various magnetic properties of a few members (\textit{R} = Gd, Tb, Ho) of \textit{R}Ir$_{3}$ series of intermetallic compounds.

\section{Experimental Details}

The polycrystalline \textit{R}Ir$_3$ (\textit{R} = Gd, Tb, Ho) compounds were synthesized in an arc-furnace by melting the stoichiometric amount of constituent elements of high purity ($>$ 99.9\%) on a water-cooled Cu hearth under flowing inert gas Ar atmosphere. The ingots were melted 5 - 6 times after flipping each time to get volume homogeneity. The weight loss after melting were less than 1\%. The as-cast ingots were annealed subsequently under vacuum in a sealed quartz tube at 900$^{\circ}$C for 7 days. The structural characterization of the annealed compounds were characterised by powder x-ray diffraction (XRD) technique at room temperature using Cu-K$\alpha$ radiation on a TTRAX-III diffractometer (M/s Rigaku, Japan) having 9 kW power supply. The crystal structure and phase purity of the annealed compounds were checked by Rietveld analysis of XRD data using FullProf software package \cite{7}. The scanning electron microscopy (SEM) Measurements were carried out in the instrument EVO 18 (M/s Carl Zeiss, Germany) and the elemental analysis were performed in energy dispersive x-ray spectroscopy (EDX)(Element EDS system, M/s EDAX Inc., USA). The dc magnetic measurements were carried out in SQUID VSM (M/s Quantum design Inc., USA) in the temperature range 2 - 300 K and magnetic fields up to 70 kOe. The ac magnetic susceptibility measurements were carried out in Ever Cool II VSM system (M/s Quantum design Inc, USA). The specific heat of the sample at zero field were measured on PPMS system (M/s Quantum design Inc, USA).\\

\section{Results and Discussions:}

\subsection{Structural analysis:}

\begin{figure}
\begin{center}
\includegraphics[scale=0.28]{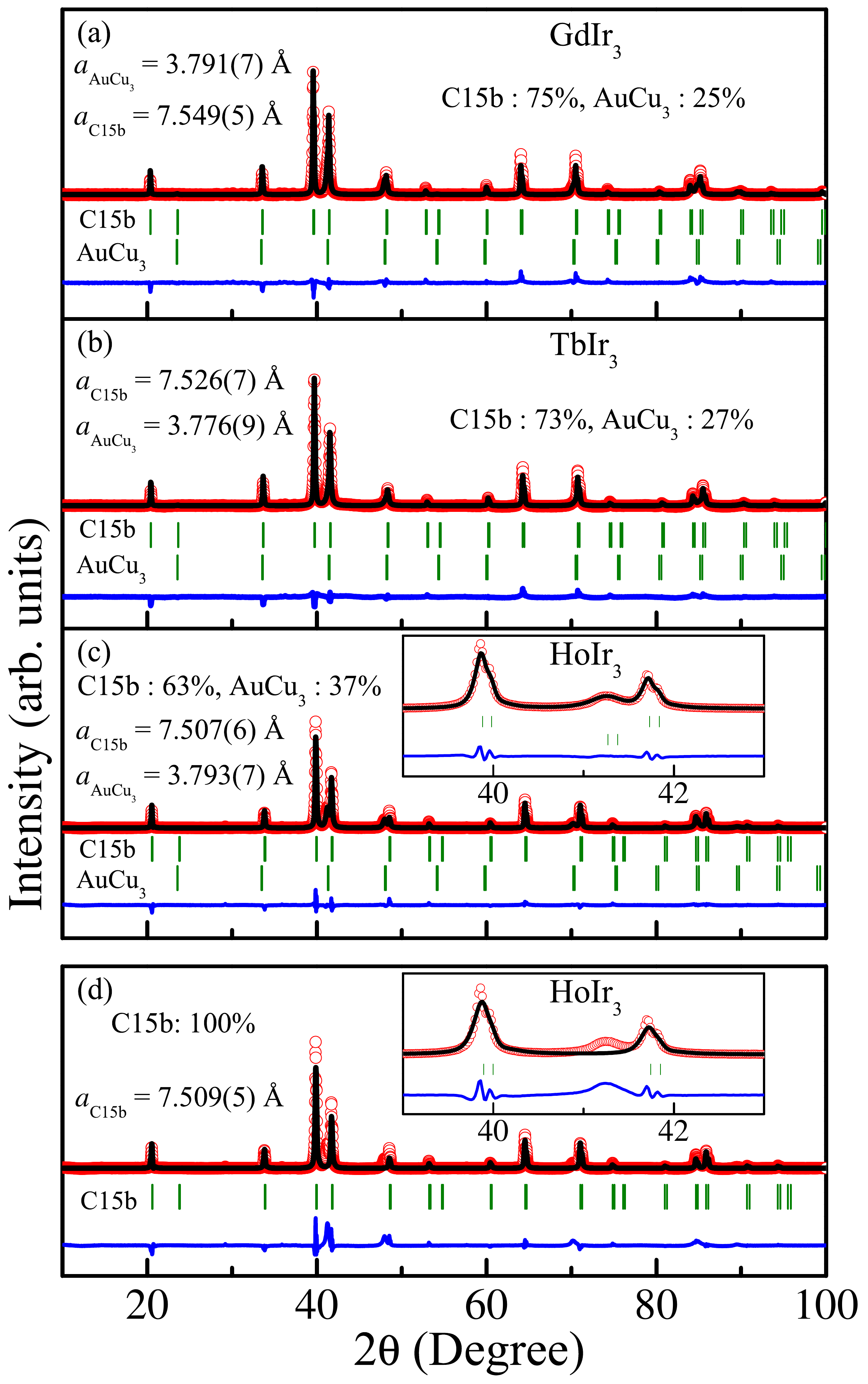}
\caption{Rietveld refinement of the powder XRD patterns at room temperature and calculated Bragg positions in (a) GdIr$_3$, (b) TbIr$_3$ and (c) HoIr$_3$ considering together C15b and AuCu$_3$-type structure and in (d) HoIr$_3$ considering only C15b-type structure. The insets in (c) and (d) show the quality of fitting in an expanded region.}
\label{fgr:XRD}
\end{center}
\end{figure}

\begin{figure}
\begin{center}
\includegraphics[scale=0.6]{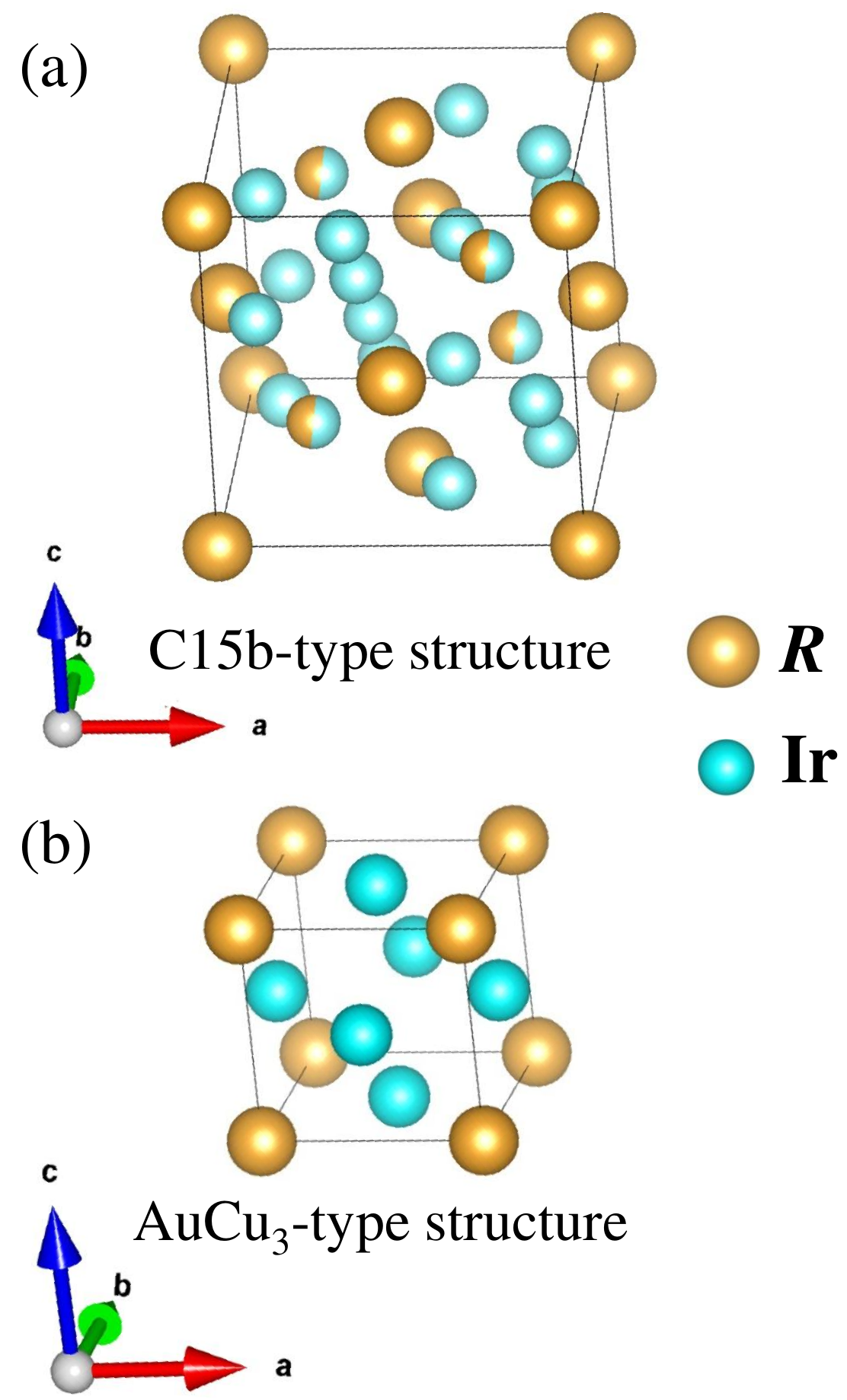}
\caption{Unit cells of (a) C15b and (b) AuCu$_3$-type structure}
\label{fgr:Unit cell}
\end{center}
\end{figure}

The powder XRD patterns of  \textit{R}Ir$_3$ (\textit{R} = Gd, Tb, Ho) compounds at room temperature are shown in fig. \ref{fgr:XRD}. The presence of sharp peaks in the XRD patterns of all samples confirm well crystalline behavior. In literature, various members of the binary \textit{AB}$_3$ system are reported to form in a wide variety of crystal (C15b, AuCu$_3$, CeNi$_3$, PuNi$_3$-type) structures depending on the rare earth and transition metals present in the system \cite{6, 8, 9, 10, 11, 12, 13}. Since the crystal structure of these \textit{R}Ir$_3$ compounds are not known \cite{13b}, we have generated XRD patterns for all possible crystal structures as mentioned above, using the PowderCell software \cite{13c}. Although the experimental XRD patterns of \textit{R}Ir$_3$ compounds closely match with the XRD pattern of cubic C15b-type structure (space group: {\it F$\bar{4}$3m}, No.216) a few additional peaks nevertheless remain unindexed for all the \textit{R}Ir$_3$ system (fig. \ref{fgr:XRD}(d)). These additional XRD peaks remain unchanged, even after the compounds were annealed at 900$^{\circ}$C for 7 days. To check the homogeneity of the materials, EDX measurement have been carried out which confirms that the \textit{R}Ir$_3$ compounds are chemically homogeneous with rare earth and transition metal ratio 1:3. It therefore appears to be quite likely that these compounds form in new crystal structure or the extra peaks may come from coexisting additional polymorphic phase similar to that reported earlier in \textit{R}Pt$_3$B series of compounds \cite{5}. It may also be pointed out here that binary TbPt$_3$ compounds are also known to form in two different polymorphic phases TbPt$_3$: C15b and AuCu$_3$-type \cite{8}. In our further analysis, we found that those additional XRD peaks of \textit{R}Ir$_3$, which remain unaccounted in C15b structure could be indexed with the cubic AuCu$_3$-type crystal structure (space group: {\it Pm$\bar{3}$m}). The detailed Rietveld analysis of the XRD patterns of all of the annealed \textit{R}Ir$_3$ (\textit{R} = Gd, Tb, Ho) compounds considering both the cubic C15b-type and cubic AuCu$_3$-type  phases are shown in fig. \ref{fgr:XRD}. In C15b-type unit cell, 4\textit{a} (0,0,0) site is occupied by \textit{R} atoms, while 16\textit{e} ($\frac{5}{8}, \frac{5}{8}, \frac{5}{8}$) site is occupied by Ir atoms. The remaining \textit{R} and Ir atoms are randomly distributed among 4\textit{c} site ($\frac{1}{4}, \frac{1}{4}, \frac{1}{4}$) in equal proportions (Fig. \ref{fgr:Unit cell}(a)). In AuCu$_3$-type unit cell the \textit{R} atoms sit in the cubic corner positions 1\textit{a} (0,0,0) while Ir atoms occupy the face centre positions 3\textit{c} (0, $\frac{1}{2}, \frac{1}{2}$) (fig. \ref{fgr:Unit cell}(b)). The details of crystallographic parameters obtained from the Rietveld refinement are listed in table \ref{table:1}. The \textit{R}Ir$_3$ system thus form as a macroscopic coexistence of two crystalline phases, cubic C15b and cubic AuCu$_3$-type with different relative percentage for different rare earths (Table \ref{table:1}). Coexistence of similar polymorphic phases have earlier been reported in literature \cite{5, 5a}. For example it is recently reported that \textit{R}Pt$_3$B compounds form in two polymorphic crystal structures, tetragonal CePt$_3$B-type and and ideal cubic perovskite-type at room temperature \cite{5}. However, after annealing at high temperature the percentage of the cubic phase increases beyond the tetragonal phase. On the other hand in \textit{R}Ir$_3$ series the percentage of two phases remain conserved upon annealing at temperature upto 900$^{\circ}$C \cite{13b} . \\

\begin{table}
\caption{Crystallographic and fitting parameters from Rietveld refinement of XRD data of \textit{R}Ir$_3$ (\textit{R} = Gd, Tb, Ho) compounds}
\label{table:1}
\centering
\vspace{0.3cm}
\begin{tabular}{lllll}
\hline \hline
Compound & Phase & {\it a}(\AA) & {\it R$_{Bragg}$} & {\it R$_f$}\\
\hline
GdIr$_3$ & C15b (74\%) & 7.5495(1) & 15.9 & 11.3\\
 & AuCu$_3$ (26\%) & 3.7917(3) & 13.6 & 19.1\\
 TbIr$_3$ & C15b (73\%) & 7.5267(1) & 9.72 & 7.70\\
 & AuCu$_3$ (27\%) & 3.7769(2) & 7.63 & 10.3\\
 HoIr$_3$ & C15b (63\%) & 7.5076(1) & 10.8 & 7.39\\
 & AuCu$_3$ (37\%) & 3.7937(2) & 5.6 & 7.77\\
 \hline \hline

\end{tabular}
\end{table}

\subsection{dc magnetization}

\begin{figure}
\begin{center}
\includegraphics[scale=0.28]{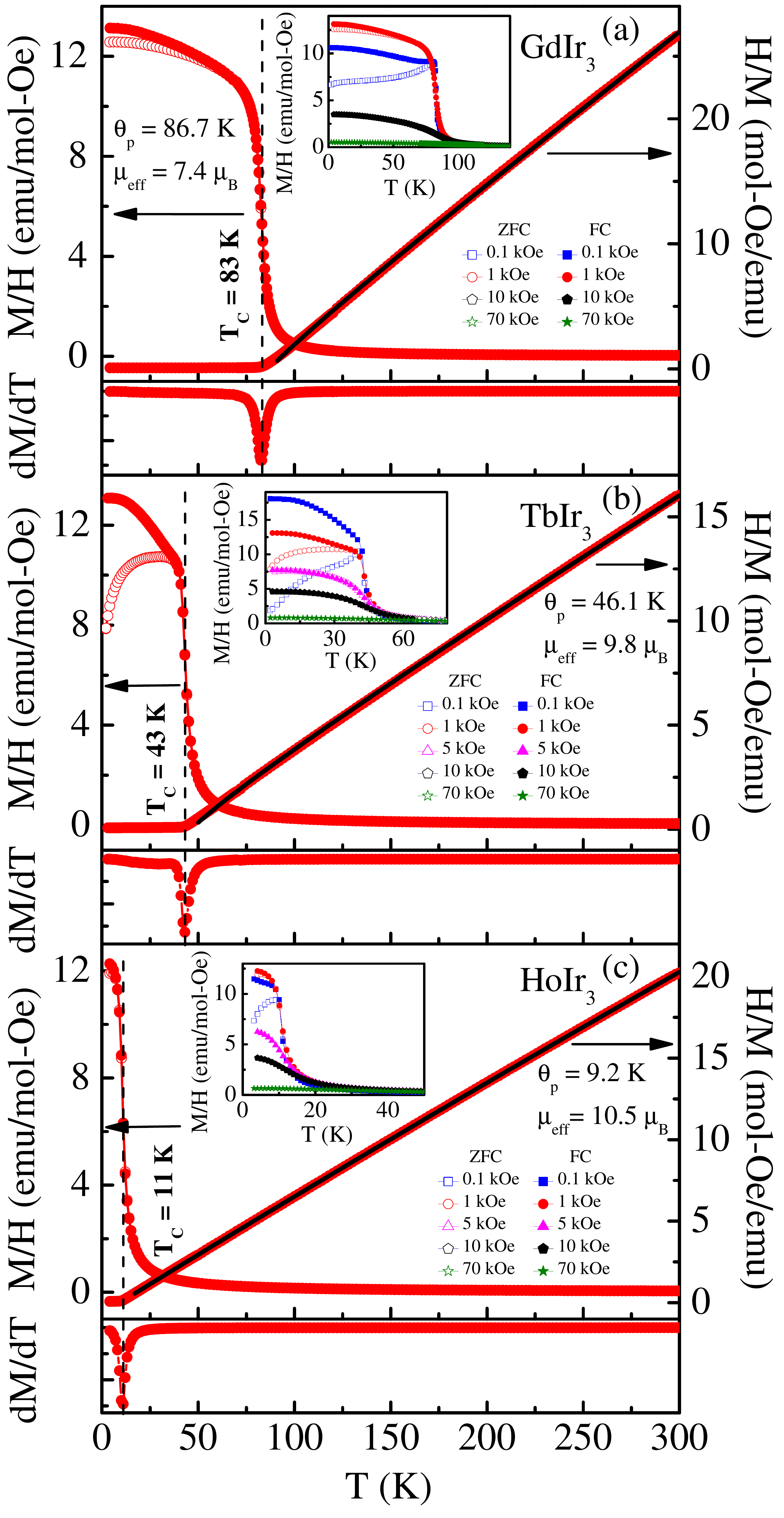}
\caption{Upper panels: Temperature dependence of dc magnetic susceptibility and inverse susceptibility in ZFC and FC mode of (a) GdIr$_3$, (b) TbIr$_3$ and (c) HoIr$_3$ compounds under applied magnetic field of 1 kOe. The solid black lines are the fit to the Curie-Weiss law. The insets show the expanded region of magnetic susceptibility below the transition temperature at different applied magnetic fields. Lower panels: Temperature derivatives of FC magnetization of (a) GdIr$_3$, (b) TbIr$_3$ and (c) HoIr$_3$ compounds under applied magnetic field of 1 kOe.}
\label{fgr:M-T}
\end{center}
\end{figure}

The temperature dependent dc magnetic susceptibilities ($\chi = M/H$) of $\textit{R}$Ir$_3$ ($\textit{R}$ = Gd, Tb, Ho) compounds under both zero-field-cooled (ZFC) and field-cooled (FC) protocols at different applied external magnetic fields are shown in fig. \ref{fgr:M-T}. The magnetic transition temperatures have been determined from the first order temperature derivative $\left(\frac{dM}{dT}\right)$ of the magnetization measured at 1 kOe applied magnetic field under FC condition. In case of ferromagnetic (FM) transition, the minima in the $\frac{dM}{dT}$ curves are described as the transition temperature $T_{\rm{C}}$ from paramagnetic to ferromagnetically ordered state. On the other hand, the transition temperature $T_{\rm{N}}$ of an antiferromagnetically ordered compound is defined as the temperature, at which the $\frac{dM}{dT}$ curve changes its sign. The temperature dependent magnetic susceptibility curves for each of these three compounds exhibit a single ferromagnetic anomaly at low temperatures. As our system consist of macroscopic coexistence of two phases (C15b-type, AuCu$_3$-type), at least one of these two phases must have undergone ferromagnetic transition at low temperatures. If we compare the magnetic properties of $\textit{R}$Ir$_3$ compounds with those of $\textit{R}$X$_3$ (X = Pd, Pt) compounds, one may able to shed some light on which phase is responsible for ferromagnetic ordering. Since the AuCu$_3$-type of structure in $\textit{R}$X$_3$ compounds is generally known to be conducive to antiferromagnetic interaction \cite{13, 13a}, the ferromagnetic interaction observed in \textit{R}Ir$_3$ compounds is likely to arise from the other polymorphic phase C15b. Absence of any antiferromagnetic signal in experimental data suggests AuCu$_3$ phase may remain paramagnetic down to lowest experimental temperature limit 2 K. This point will be discussed further later in this work, where quantitative analysis of magnetic entropy have been carried out.\\

The \textit{R}Ir$_3$ series of compounds show ferromagnetic transition at low temperatures with $T_{\rm{C}}$ = 83 K, 43 K, 11 K, respectively for GdIr$_3$, TbIr$_3$ and HoIr$_3$. Magnetic susceptibility measurements of these compounds reveal thermal hysteresis behavior below their respective ordering temperatures. As the strength of the applied magnetic field increases, the same thermal hysteresis tend to weaken gradually. The temperature below which the divergence appears also decreases with the increasing applied magnetic field. At high applied magnetic field \textit{e.g.} at 70 kOe, we obtain a completely reversible nature of temperature dependence of magnetic susceptibility. Such a thermal irreversibility between the ZFC and FC magnetization are generally attributed to the magnetic anisotropy and/or /spin/cluster glass behavior present in this system. We will probe on further detail on this point while discussing the magnetic relaxation phenomenon.\\

The observed magnetic transition temperatures of these compounds found to get reduced with increasing atomic number of rare earth element in the \textit{R}Ir$_3$ series of materials (\textit{R} = Gd and heavier rare earth). Generally, such a reduction of magnetic ordering temperature can be explained using de-Gennes scaling behavior of isostructural series of compounds. However, we found that the measured magnetic ordering temperatures deviates strongly than those expected from the de-Gennes scaling with respect to the transition temperature of GdIr$_3$ where the crystalline electric field effect is negligible. Conventionally, a good agreement between the experimental value of transition temperatures and that obtained from the de-Gennes scaling is an indication of the dominance of RKKY interaction over the crystalline electric field (CEF) effect. The discrepancy between the experimental and the scaled value may therefore indicates that the CEF level scheme might have a strong influence on the magnetic ordering temperature in these materials. Inelastic neutron scattering experiment may help us to identify the CEF level scheme of the rare earth ions in these materials. However, these measurements are beyond the scope of the present work.\\

In the paramagnetic region $T_{\rm{C}}$ $<$ T $\leq$ 300 K magnetic susceptibility curves follow the Curie-Weiss behavior:

\begin{equation}
\chi = \chi_0 + \frac{C}{(T - \theta_p)}
\end{equation}

\noindent where C is the Curie constant, $\theta_p$ is the paramagnetic Curie temperature and $\chi$$_0$ is the temperature independent contribution. The estimated values of effective magnetic moment ($\mu$$_{eff}$) and paramagnetic Curie temperature ($\theta_p$) are mentioned in fig. \ref{fgr:M-T}. The $\mu_{eff}$ values for \textit{R}Ir$_3$ compounds closely follow those of the theoretical free ion values $\sqrt{gJ(J+1)}$ of the respective \textit{R}$^{3+}$ ion, indicating that only the localized 4{\it f} shells of \textit{R}$^{3+}$ are contributing towards the magnetism. The estimated $\theta_p$ values appear to be very close to the experimental value of $T_{\rm{C}}$, as expected in most of the ferromagnetic materials.\\

\subsection{Heat Capacity}

\begin{figure}
\begin{center}
\includegraphics[scale=0.28]{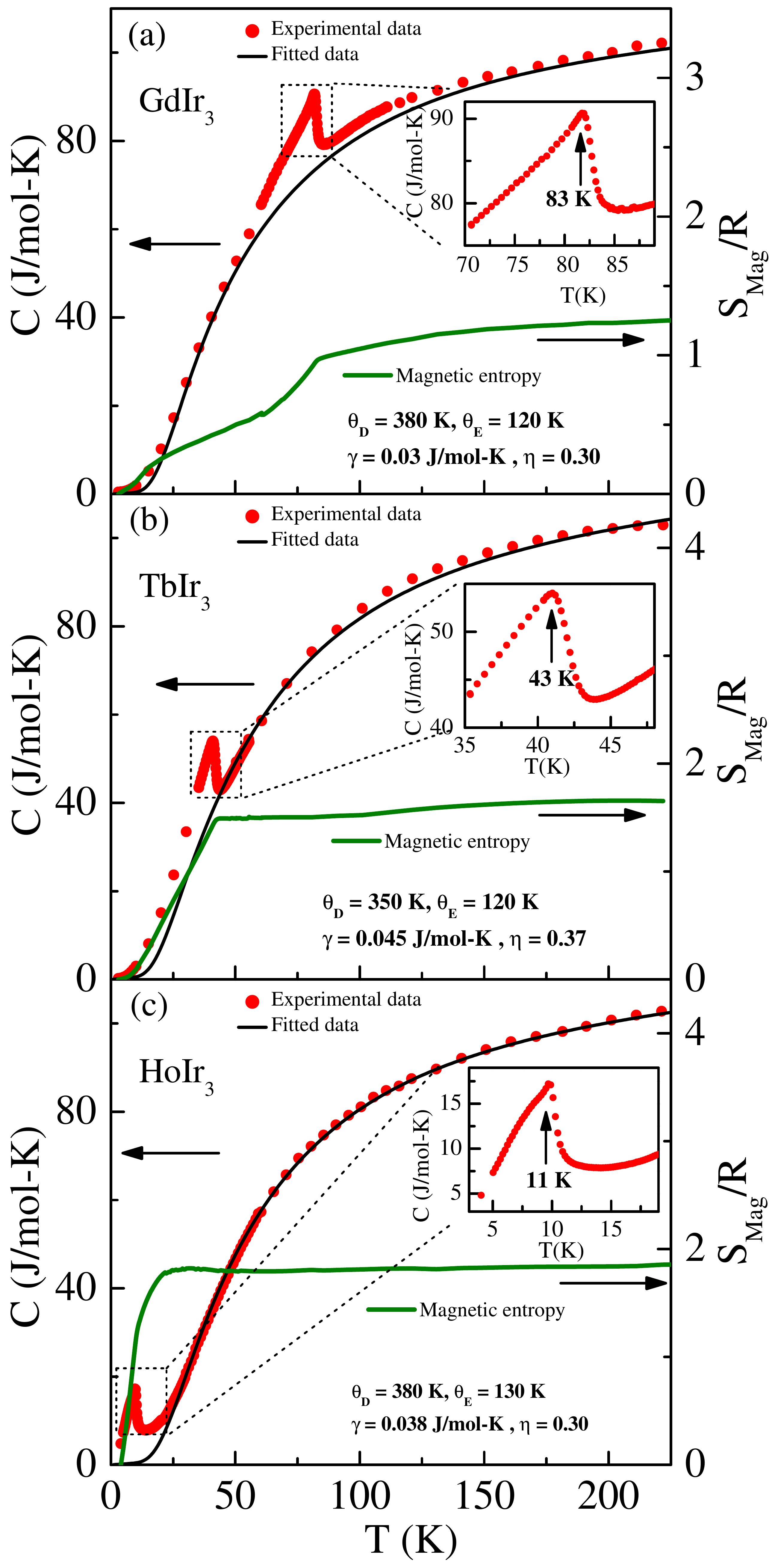}
\caption{The temperature dependence of the observed and fitted heat capacity data in absence of any magnetic field and the calculated magnetic entropy of (a) GdIr$_3$ (b) TbIr$_3$ and HoIr$_3$ compounds. The insets show the zero-field heat capacity in an expanded region around their respective transition temperature.}
\label{fgr:Sp heat}
\end{center}
\end{figure}

As mentioned earlier, heat capacity measurement is generally considered to be an effective tool to establish the bulk nature of magnetic ordering. An estimation of magnetic entropy from the heat capacity measurements can also provide valuable information regarding the volume fraction involved in the magnetic ordering process. Fig. \ref{fgr:Sp heat} shows the heat capacity data of \textit{R}Ir$_3$ (\textit{R} = Gd, Tb, Ho) compounds as a function of temperature at zero applied magnetic field. A $\lambda$-like transition is observed around 83 K, 43 K and 11 K respectively for GdIr$_3$, TbIr$_3$, HoIr$_3$ that are in good agreement with the magnetization measurements. The single $\lambda$-like transition observed in the heat capacity measurement corresponds to the long range magnetic ordering of the respective \textit{R}Ir$_3$ compounds.\\

The temperature dependent heat capacity can be described by the standard formula,

\begin{equation}
C_p = \gamma T + C_{phonon} + C_{mag}
\end{equation}

\noindent Here the first, second and third terms correspond to the electronic, phononic and magnetic contributions, respectively. At high temperature the heat capacity approaches to the classical value 3NR ($\sim$100 J/mol-K for N = 4 in case of \textit{R}Ir$_3$). The magnetic contribution to heat capacity is generally estimated by subtracting the heat capacity of isostructural non-magnetic analogue from the heat capacity of the magnetic member, as it is generally assumed that the lattice and electronic contribution of both the system remain essentially same. However since in our case the system effectively consists of two different polymorphic phases where effective volume fraction depends strongly on the rare earth members involved, the same standard procedure may turn out to be quite misleading. Insted, one may first estimate the electronic and phononic contribution to heat capacity by fitting the data in the paramagnetic region and then extrapolated the fit down to 0 K. By subtracting the fitted curve from the experimental data, the $C_{mag}$ can subsequently be determined.\\

The total heat capacity of a material in the paramagnetic region consists of two contributions: electronic ($\gamma$T) and Phononic ($C_{phonon}$). The phononic contribution was first explained by Einstein, who assumed that a solid composed of N atoms can be represented as 3N independent harmonic oscillators having same frequency \cite{16, 16a}. The Einstein contribution can be written as \cite{16, 17},

\begin{equation}
C{\rm_E} = \sum_N 3N{\rm_ER}\frac{x^2{\rm e}^x}{[{\rm e}^x-1]^2},
\end{equation}

\noindent where $N{\rm_E}$ is the number of Einstein oscillators, \textit{x} = $\theta{\rm_E}/T$, $\theta{\rm_E}$ is the Einstein temperature. However, it was found that the Einstein model appears to be quite inadequate to describe the experimentally observed specific heat behavior at low temperature region for most of the solids \cite{16, 16a, 16b}. Following this, Debye had modified Einstein model by assuming that the solid consisting of a set of coupled oscillator instead of independent oscillators \cite{16, 16a, 16b}, where the phononic contribution to heat capacity takes the following form \cite{16, 17},

\begin{equation}
C_{\rm{D}} = 9N_{\rm{D}}{\rm R}\left(\frac{T}{\theta_{\rm{D}}}\right)^3\int_0^{\frac{\theta_{\rm{D}}}{T}}\frac{x^2{\rm e}^xdx}{[{\rm e}^x-1]^2}
\end{equation}

\noindent where $N\rm_D$ is the number of Debye oscillators and \textit{x} = $\theta{\rm_D}/T$, $\theta{\rm_D}$ being the Debye temperature. The modification proposed by Debye indeed able to explain the low temperature heat capacity data in much better way than Einstein model. The Debye model still cannot describe the experimental heat capacity behavior over the entire temperature region, as it works well below $\theta{\rm_D}/50$ and above $\theta{\rm_D}/10$ only \cite{16}. The quantitative mismatch in the intermediate temperature region has its origin in the fact that the phonon dispersion phenomenon was not taken into account in the Debye model. Since neither a single Einstein model nor a single Debye model can describe the experimental outcome over the whole temperature range, a combination of both the contributions generally used to describe the overall heat capacity behavior \cite{16c, 16d, 16e, 16f, 16g, 16h, 17}, that can be expressed as \cite{16g, 16h, 17},

\begin{equation}
C_{P} = \gamma T + \sum_i\eta_{i}C_{Ei}(T)+\left(1-\sum_i\eta_{i}\right)C_{D}(T)\label{eqn:HC4}
\end{equation}

\noindent The parameter $\eta$ determines the relative percentage of the two contributions. We have achieved a good fit for all these three compounds in the paramagnetic region by using eqn. (\ref{eqn:HC4}) with Einstein and Debye contributions for GdIr$_3$, TbIr$_3$ and HoIr$_3$ as 70\% and 30\%, 63\% and 37\%, 70\%, 30\% respectively. The corresponding fitting parameters are listed in fig.\ref{fgr:Sp heat}. $C_{mag}$ has been evaluated by subtracting the experimental data from the fitted model after extrapolating to lowest temperature. After calculating $C_{mag}$ and integrating $\frac{C_{mag}}{T}$, over the entire temperature range, it is possible to estimate the magnetic entropy, $S_{mag}$ (= $\int_{0}^{T} \frac{C_{mag}}{T} dT$). For a bulk magnetic phenomenon, when all the \textit{R} ions takes part in the magnetic ordering process, at high temperature, $S_{mag}$ saturates to the theoretical value R$ln(2J+1)$, where \textit{J} is the total angular momentum and R (= 8.31 J/K) is the universal gas constant. Theoretically if all the \textit{R} atoms would have ordered, one would expect $S_{mag}$ to be reached to R\textit{ln}8, R\textit{ln}13 and R\textit{ln}17 for GdIr$_3$ (\textit{J} = 7/2), TbIr$_3$ (\textit{J} = 6), HoIr$_3$ (\textit{J} = 8) respectively. In our analysis, we however found a much reduced value of $S_{mag}$ as 1.3R for GdIr$_3$, 1.7R for TbIr$_3$ and 1.7R for HoIr$_3$. Thus we have found from magnetic entropy calculation that only 70\% of Gd atom, 70\% of Tb atom and 62\% of Ho atom  would have contribute towards magnetism. Since in XRD it was found that \textit{R}Ir$_3$ (\textit{R} = Gd, Tb and Ho) consists of two different phases \textit{viz}, C15b and AuCu$_3$-type with relative percentage 74 and 26, 73 and 27, 63 and 37 respectively, the reduced value of estimated $S_{mag}$ indicates that only the C15b phase participate in magnetism.\\

\begin{figure}
\begin{center}
\includegraphics[scale=0.28]{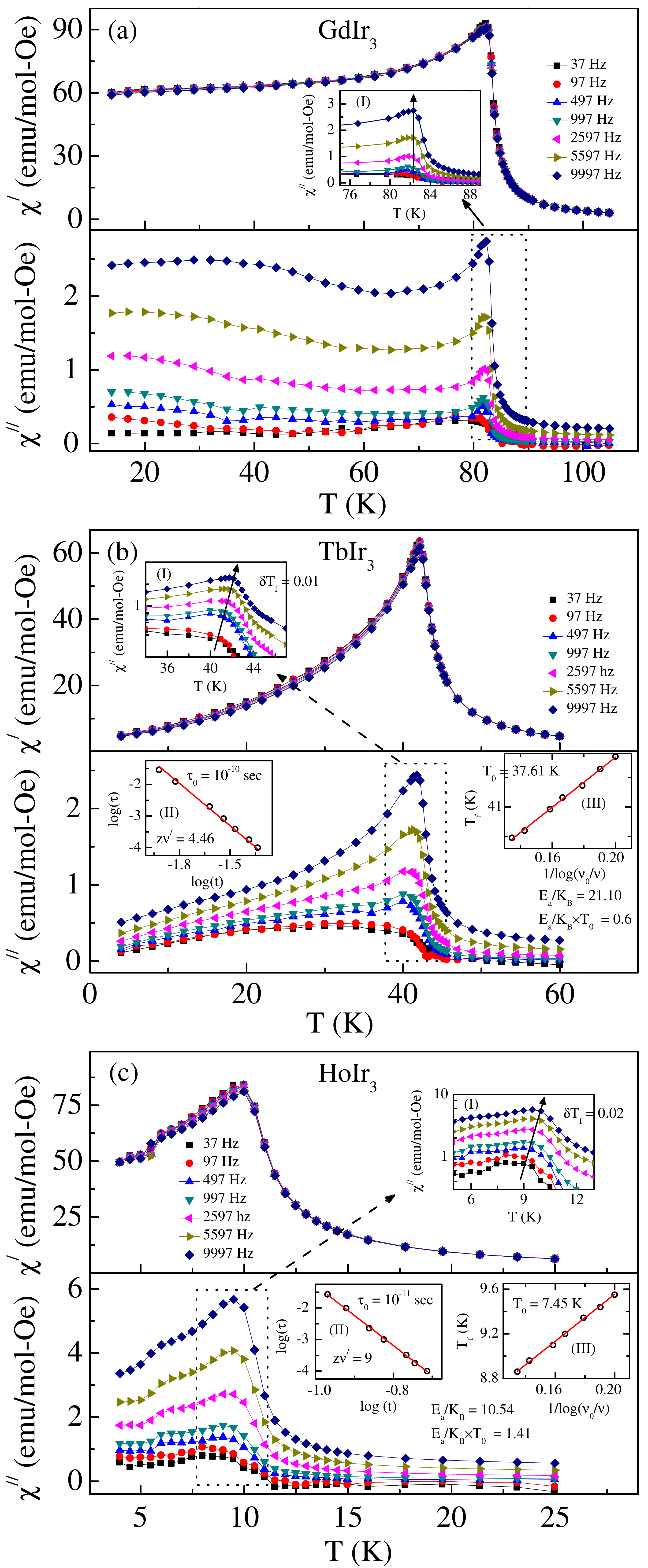}
\caption{The temperature dependent of real (upper panel) and imaginary part (lower panel) of ac magnetic susceptibility of (a) GdIr$_3$, (b) TbIr$_3$ and (c) HoIr$_3$ compounds at different frequencies. The insets in the upper panel show the expanded region near the transition temperature for all the three compounds. The left insets in the lower panel of (b) TbIr$_3$ and (c) HoIr$_3$ show the variation of ln($\tau$)with ln(t). The solid lines represent the fit to the power-law divergence. The frequency dependence of the transition temperature is shown in the right insets for (b) TbIr$_3$ and (c) HoIr$_3$ compounds. The solid lines represent the fit to the Vogel-Fulcher law.}
\label{fgr:AC susceptibility}
\end{center}
\end{figure}
\subsection{ac succeptibility}

The ac susceptibility measurements were carried out for GdIr$_3$, TbIr$_3$ and HoIr$_3$ in an excited field of 0.124 Oe for different frequencies (\textit{f}). Fig. \ref{fgr:AC susceptibility} shows the variations of real and imaginary parts of magnetic susceptibility with temperature at various frequencies. For all the three samples, the peaks in $\chi^{\prime}$ as well as $\chi^{\prime \prime}$ could be found at the respective temperatures, which have been identified as magnetic ordering temperatures through dc magnetic susceptibility as well as heat capacity measurements. The non-zero values of $\chi^{\prime \prime}$ below the ordering temperatures suggest the magnetic ordering to be ferromagnetic type.\\

The peak temperature in both $\chi^{\prime}$ and $\chi^{\prime \prime}$ in GdIr$_{3}$ remain invariant as a function of frequency indicating a long range nature of the ferromagnetic ordering in this compound. However, although the peak temperatures in $\chi^{\prime}$ for TbIr$_{3}$ and HoIr$_{3}$ do not exhibit any discernible shift as a function of frequency, a close examination of $\chi^{\prime \prime}$ for both the samples reveal a change in peak position. As the frequency increases, the peak in $\chi^{\prime \prime}$ tend to shift toward higher temperature. This feature is generally attributed to the presence of metastable spins, in the system that exhibit spin/cluster glass behavior. For TbIr$_3$ the peak temperature shifts from 40.50 K to 41.92 K with increasing frequency from 37 Hz to 9997 Hz [fig. \ref{fgr:AC susceptibility} (b)]. In case of HoIr$_3$, the peak shifts from 9 K to 9.6 K with similar increase in frequency [fig. \ref{fgr:AC susceptibility} (c)].\\

As mentioned above, such a shift in peak temperature manifests the presence of spin/cluster glass transition with $T_{f}$ (freezing temperature) to be 40 K for TbIr$_3$ and 9 K for HoIr$_3$. To find the catagory of spin/cluster glass system we have estimated the relative freezing temperature shift per decade of frequency which is defined as \cite{18}.\\

\begin{eqnarray}
\delta T_{f} =\frac{\Delta T_{f}}{T_{f}\Delta (log_{10}\nu)}
\end{eqnarray}

\noindent where $T_{f}$ is the freezing temperature, $\nu$ is the applied frequency. The value of $\delta T_{f}$ found to be 0.01 for TbIr$_3$ and 0.02 for HoIr$_3$, which suggest the systems belonging to canonical spin glass type \cite{18}.\\

The spin glass nature can also be established through the power law behavior of $T_{f}$ that \cite{18,19}

\begin{eqnarray}
\tau = \tau_0\left(\frac{T_f-T_{SG}}{T_{SG}}\right)^{-z\nu^{\prime}}
\end{eqnarray}

\noindent where $\tau = \frac{1}{\nu}$ is the relaxation time corresponding to the applied frequency, $\tau_0$ is the relaxation time for single spin-flip, $T_{SG}$ is the temperature of spin/cluster glass with f = 0, $\nu^{\prime}$ is known as critical exponent for correlation length $\xi = \left(\frac{T_f}{T_{SG}}-1\right)^{-\nu{\prime}}$ and $\tau \sim\xi^z$. The term $z\nu^{\prime}$ is called as dynamical critical exponent. For canonical spin glass, the value of critical exponent $z\nu^{\prime}$ lies between 4 and 12 while $\tau_0$ lies between $10^{-12}-10^{-13}$ \cite{20}. In our analysis, we have estimated the values of $z\nu^{\prime}$ \& $\tau_0$ to be 4.46 \& $10^{-10}$ sec for TbIr$_3$ and 9, $10^{-11}$ sec for HoIr$_3$ respectively. The derived values of $z\nu^{\prime}$ are in the range reported for spin-glass system but $\tau_0$ value is somewhat larger than canonical spin glass although it remains orders of magnitude larger than the values for cluster glass system ($\tau_0 \sim 10^{-7}$) \cite{20, 20a, 20b, 20c, 20d}.\\

Vogel-Fulcher relation \cite{18, 21} is another dynamical scaling law for spin/cluster glass system where freezing temperature $T_f$ depends on frequency as

\begin{eqnarray}
\nu = \nu_0{\rm exp}\left[-\frac{E_a}{k_B(T_f-T_0)}\right]
\end{eqnarray}
\noindent Here $E_a$ denotes the activation energy, $\nu_0$ the characteristic attempt frequency and $T_0$ the Vogel-Fulcher temperature. For typical canonical spin-glass, $\frac{E_a}{k_B}/T_0$ value should be close to 1. We found the values of $\frac{E_a}{k_B}$, $T_0$ and $\frac{E_a}{k_B}/T_0$ as 21.10, 37.61 K and 0.6 for TbIr$_3$ and 10.54, 7.45 K and 1.41 for HoIr$_3$. These values again establish these two compounds to be canonical spin glass type.\\

\begin{figure}
\begin{center}
\includegraphics[scale=0.38]{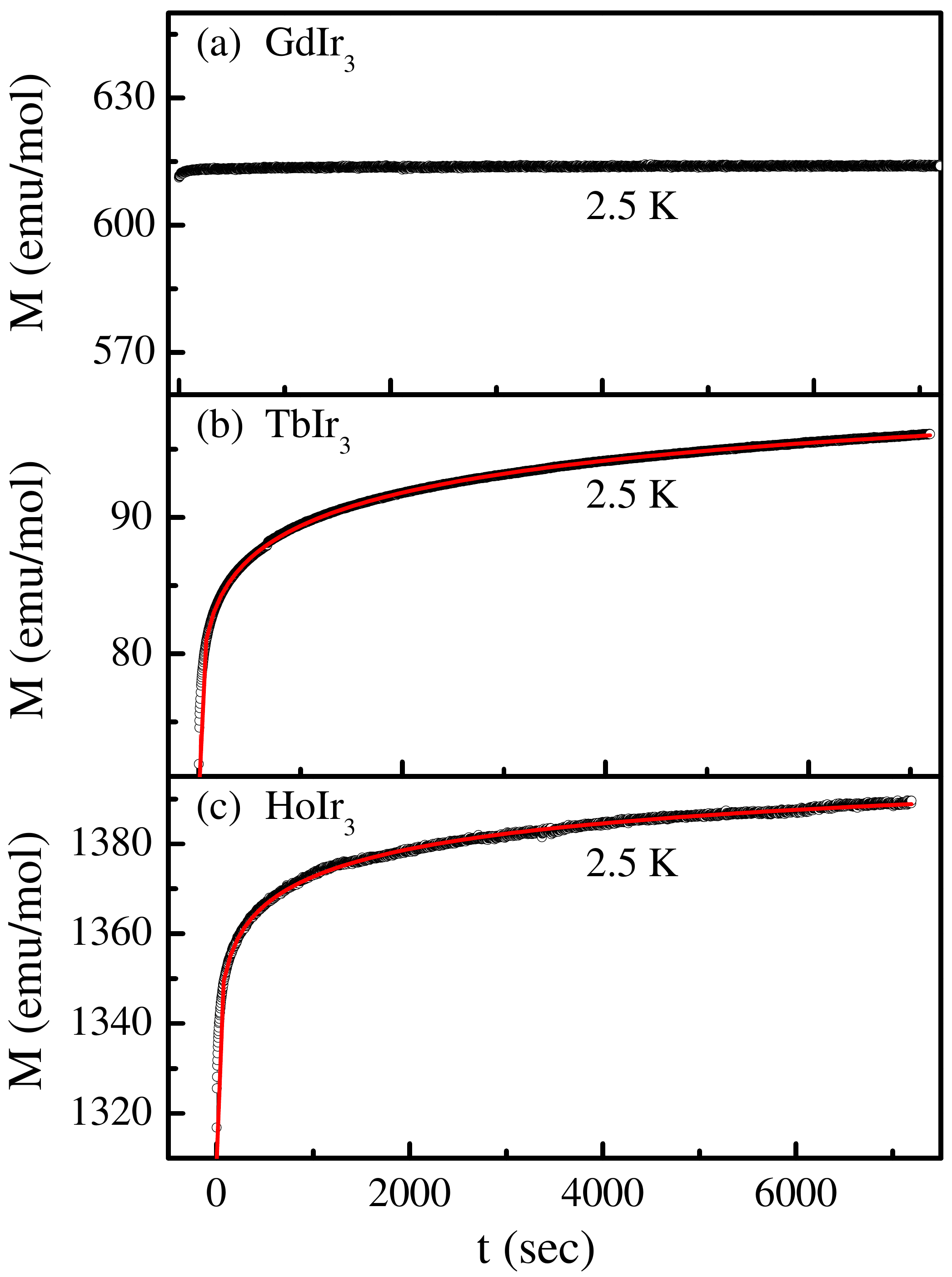}
\caption{Relaxation of Zero-field cooled magnetization at T = 2.5 K for (a) GdIr$_3$, (b) TbIr$_3$ and (c) HoIr$_3$ compounds . Solid red lines are the fit to the eqn. \ref{eqn:Stretching} }
\label{fgr:Relaxation}
\end{center}
\end{figure}
\subsection{Nonequilibrium dynamics}

The presence of magnetically frustrated spins in the system can also be established by studying the magnetic relaxation behaviors. The relaxation process have been studied under ZFC protocol, where the sample is cooled in the absence of any external applied field from the paramagnetic region to the desired temperature which is below $T_{\rm{C}}$. After reaching the desired temperature, the sample is kept at zero field for a certain time  at that temperature. Subsequently a small amount of field is applied and time evolution of magnetization (M(t)) is monitored. The ZFC relaxation of \textit{R}Ir$_3$ (\textit{R} = Gd, Tb, Ho) compounds at temperature 2.5 K are displayed in the fig.\ref{fgr:Relaxation}. As expected earlier, GdIr$_{3}$ which does not show glassy behavior, exhibit no magnetic relaxation [\ref{fgr:AC susceptibility}(a)].\\

The time dependent magnetization follow the exponential behavior as \cite{22, 23, 25}

\begin{equation}
\label{eqn:Stretching}
M(t) = M_0 + M_g{\rm exp}\left[-\left(\frac{t}{\tau}\right)^\beta\right]
\end{equation}

\noindent where $M_0$ is intrinsic magnetization, $M_g$ is the glassy component of magnetization, $\tau$ is the relaxation time, $\beta$ is the stretching exponent. The value of $\beta$ depends on the nature of energy barriers involves in the relaxation process. $\beta$ = 0 implies no relaxation and $\beta$ = 1 is for single time constant relaxation process. Since typical spin glass systems are characterized with a distribution of energy barriers, value of $\beta$ lies between 0 and 1 \cite{18, 24}.\\

The time evolution of magnetization of TbIr$_{3}$ and HoIr$_{3}$ are fitted with eq. (\ref{eqn:Stretching}). For compounds TbIr$_3$ and HoIr$_3$ the relaxation times obtained are 1244 sec, 600 sec and $\beta$ values have been estimated to be 0.29, 0.27 respectively. The $\tau$ value of TbIr$_{3}$ lies within the range of earlier reported different glassy systems \cite{18, 24} but $\tau$ value in case of HoIr$_{3}$ indicates a weaker nature of glassy behavior.

\begin{figure}
\begin{center}
\includegraphics[scale=0.72]{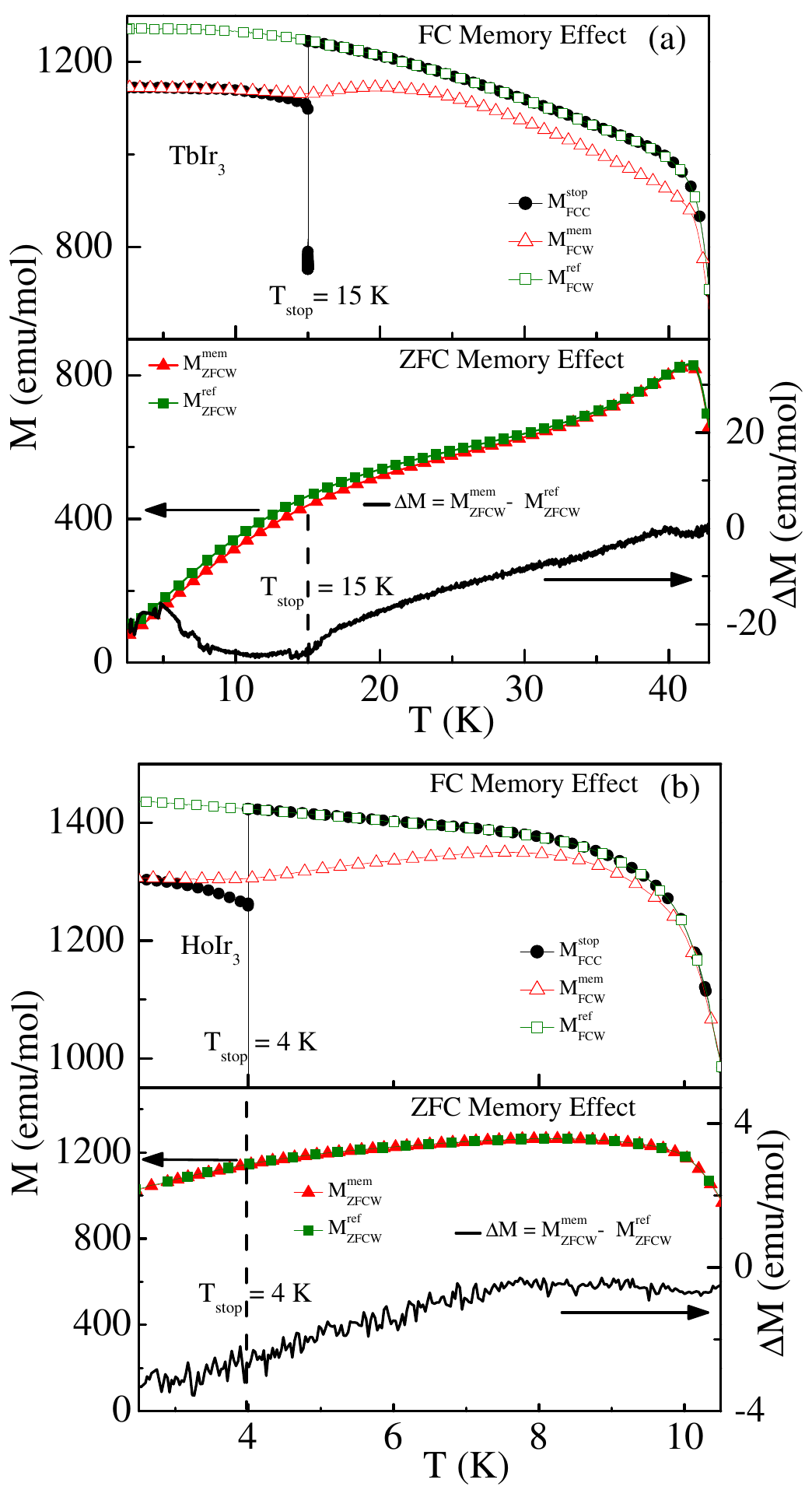}
\caption{Memory effects in the FC (upper panel) and ZFC (lower panel) protocol of (a) TbIr$_3$ and (c) HoIr$_3$ compounds. The details are discussed in the text.}
\label{fgr:Memory}
\end{center}
\end{figure}

\subsection{Magnetic memory effect}

In the foregoing part it has been observed that TbIr$_3$ and HoIr$_3$ show time dependent magnetic relaxation behavior which is absent in GdIr$_3$. Besides the magnetic relaxation behavior, magnetic memory effect is another tool for distinguishing various class of glassy systems. \\

The memory effect in \textit{R}Ir$_3$ (\textit{R} = Tb, Ho) samples have been investigated in both FC and ZFC protocols \cite{26}. In the FC protocol, the samples were cooled with low applied field of 100 Oe from the paramagnetic region (T = 125 K for TbIr$_3$ and T = 60 K for HoIr$_3$) to the lowest measurable temperature 2.5 K with single intermediate stop at T$_{stop}$ = 15 K (for TbIr$_3$) and at T$_{stop}$ = 4 K (for HoIr$_3$) for a duration of $t_W$ = 1.5 h. The magnetizations measured during this process is represented as $M^{STOP}_{FCC}$ as shown in fig. \ref{fgr:Memory}.

At the respective stopping temperatures of the two samples, the magnetic field was switched off and after the lapse of time $t_W$ = 1.5 h, the same field was reapplied with resumed cooling. After reaching lowest temperature 2.5 K, the samples are heated up to the paramagnetic region with the same applied field, as well as same rate and the measured magnetization curve is depicted as $M^{mem}_{FCW}$. The $M^{mem}_{FCW}$ curve thus obtained show a tendency to follow $M^{STOP}_{FCC}$ curve yielding a signature to remember the past history [fig. \ref{fgr:Memory}]. The standard FC magnetization curve $M^{ref}_{FCW}$ for both compounds are also displayed in the fig. \ref{fgr:Memory} (a), (b). From the figures it is clear that in case of TbIr$_3$, the observed memory effect below $T_{\rm{C}}$ is relatively stronger than that observed in case of HoIr$_3$. Such type of memory effects are quite well known behavior observed in various glassy systems \cite{23, 27}. In this context it should be mentioned that the estimated relaxation time constant estimated for HoIr$_3$ is smaller than that of TbIr$_3$.\\

The memory effect under ZFC protocol is also carried out in both these compounds. In the ZFC protocol the samples were first cooled down in zero field from the paramagnetic region to some stopping temperatures (T$_{stop}$ = 15 K for TbIr$_3$ and T$_{stop}$ = 4 K for HoIr$_3$) where the temperatures were kept on hold for t$_W$ = 1.5 h. The cooling was then recommenced down to the lowest temperature 2.5 K. The magnetization \textit{M(T)} was then recorded during heating from 2.5K to paramagnetic region under application of 100 Oe magnetic field. The \textit{M(T)} curve obtained in this process is leveled as $M^{mem}_{ZFCW}$. The reference ZFC magnetization for 100 Oe field is also measured. This is indicated as $M^{ref}_{ZFCW}$. The ZFC memory effect of these two samples are shown in the fig. \ref{fgr:Memory}. Fig \ref{fgr:Memory} also show the difference in magnetization of the two measurement processes of \textit{R}Ir$_3$ samples. The difference, $\Delta M$ (= $M^{mem}_{ZFCW}$- $M^{ref}_{ZFCW}$) shows memory dip around the stopping temperatures for both the samples.\\

It may be pointed out here that the memory effect is also observed in phase-separated or superparamagnetic systems in FC process \cite{28}. Only the ZFC memory effect can differentiate spin glass class from superparamagnetic system because superparamagnetic compound does not show memory effect in ZFC protocol \cite{28}. Thus the observed memory effect in ZFC mode confirm the presence of spin glass state in the two compounds.\\

\begin{figure*}
\begin{center}
\includegraphics[scale=1.0]{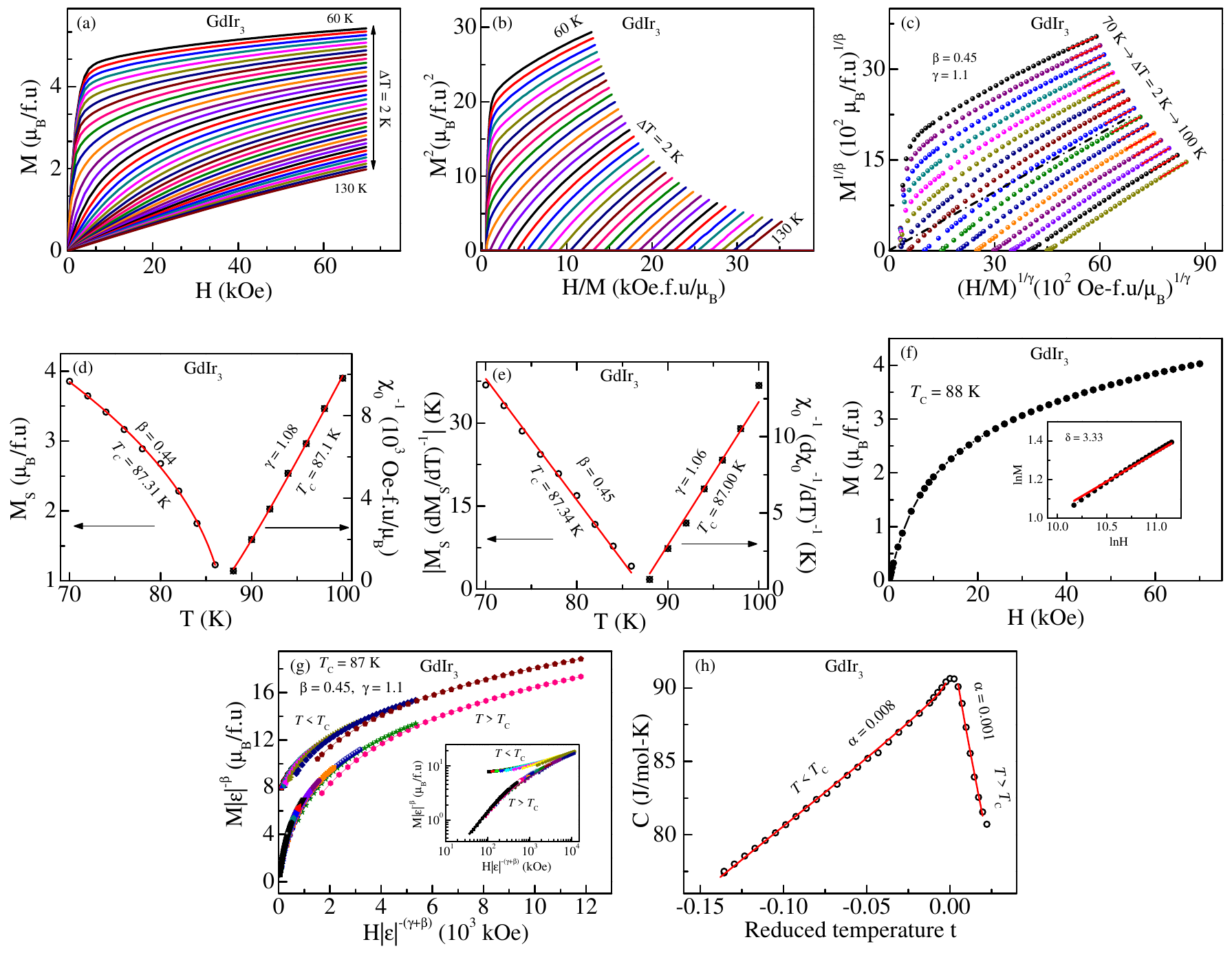}
\caption{Critical behavior of GdIr$_{3}$: (a) Isothermal magnetization curves at temperatures around ($T_{\rm{C}}$). (b) Arrott plot at different temperatures close to the Curie temperature ($T_{\rm{C}}$); (c) Modified Arrott plot. Solid lines are the linear fit of the isotherms at high field region. The isotherm close to the Curie temperature ( $T_{\rm{C}} \sim$ 87 K) almost passes through the origin; (d) Temperature dependence of spontaneous magnetization and inverse initial susceptibility. The solid lines are the fit to the power law eqs. (\ref{eqn:powerlaw1}), (\ref{eqn:powerlaw2}); (e) Kouvel-Fisher plot of spontaneous magnetization and inverse initial susceptibility. Solid lines are the linear fit to the data; (f) Critical isotherm close to the Curie temperature ($T_{\rm{C}}$). The inset shows the same on log-log scale. The solid line is the linear fit following eq. (\ref{eqn:powerlaw3}); (g) Scaled magnetization below and above  $T_{\rm{C}}$. This plot shows that all the data collapse onto two different curves: one below $T_{\rm{C}}$ and another above  $T_{\rm{C}}$. Inset shows the same on a log scale; (h) Heat capacity data on a reduced temperature scale below and above  $T_{\rm{C}}$. The solid line is the linear fit following eq. (\ref{eqn:Critical_Spheat}).}
\label{fgr:Critical_GdIr3}
\end{center}
\end{figure*}

\begin{figure*}
\begin{center}
\includegraphics[scale=1.0]{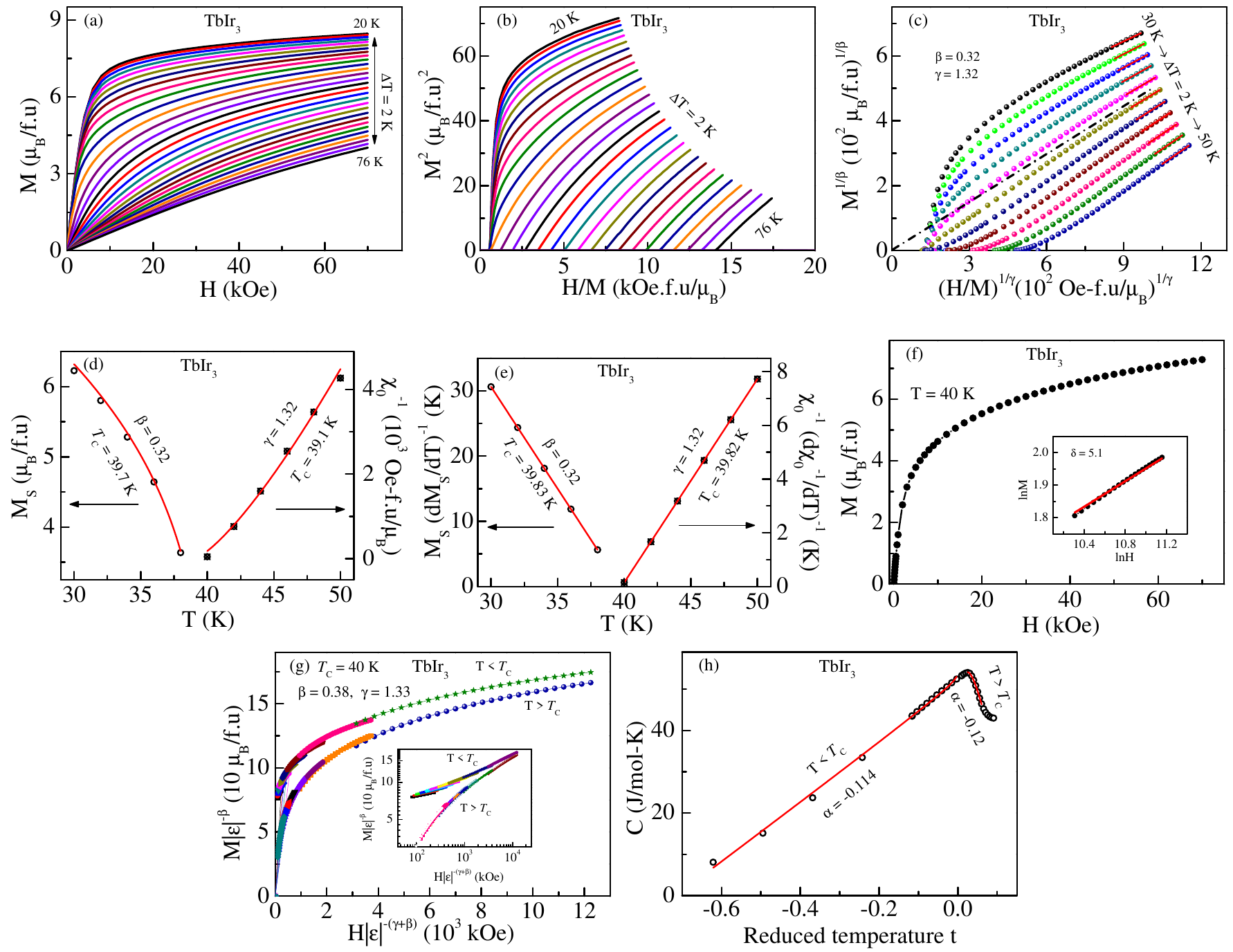}
\caption{Critical behavior of TbIr$_{3}$: (a) Isothermal magnetization curves at temperatures around ($T_{\rm{C}}$); (b) Arrott plot at different temperatures close to the Curie temperature ($T_{\rm{C}}$); (c) Modified Arrott plot. Solid lines are the linear fit of the isotherms at high temperature region. The isotherm close to the Curie temperature ($T_{\rm{C}} \sim$  40 K) almost passes through the origin; (d) Temperature dependence of spontaneous magnetization and inverse initial susceptibility. The solid lines are the fit to the power law eqs. (\ref{eqn:powerlaw1}), (\ref{eqn:powerlaw2}); (e) Kouvel-Fisher plot of spontaneous magnetization and inverse initial susceptibility. Solid lines are the linear fit to the data; (f) Critical isotherm close to the Curie temperature ($T_{\rm{C}}$). The inset shows the same on log-log scale. The solid line is the linear fit following eq. (\ref{eqn:powerlaw3}); (g) Scaled magnetization below and above  $T_{\rm{C}}$. This plot shows that all the data collapse onto two different curves: one below $T_{\rm{C}}$ and another above  $T_{\rm{C}}$. Inset shows the same on a log scale; (h) Heat capacity data on a reduced temperature scale below and above  $T_{\rm{C}}$. The solid line is the linear fit following eq. (\ref{eqn:Critical_Spheat}).}
\label{fgr:Critical_TbIr3}
\end{center}
\end{figure*}

\begin{figure*}
\begin{center}
\includegraphics[scale=1.0]{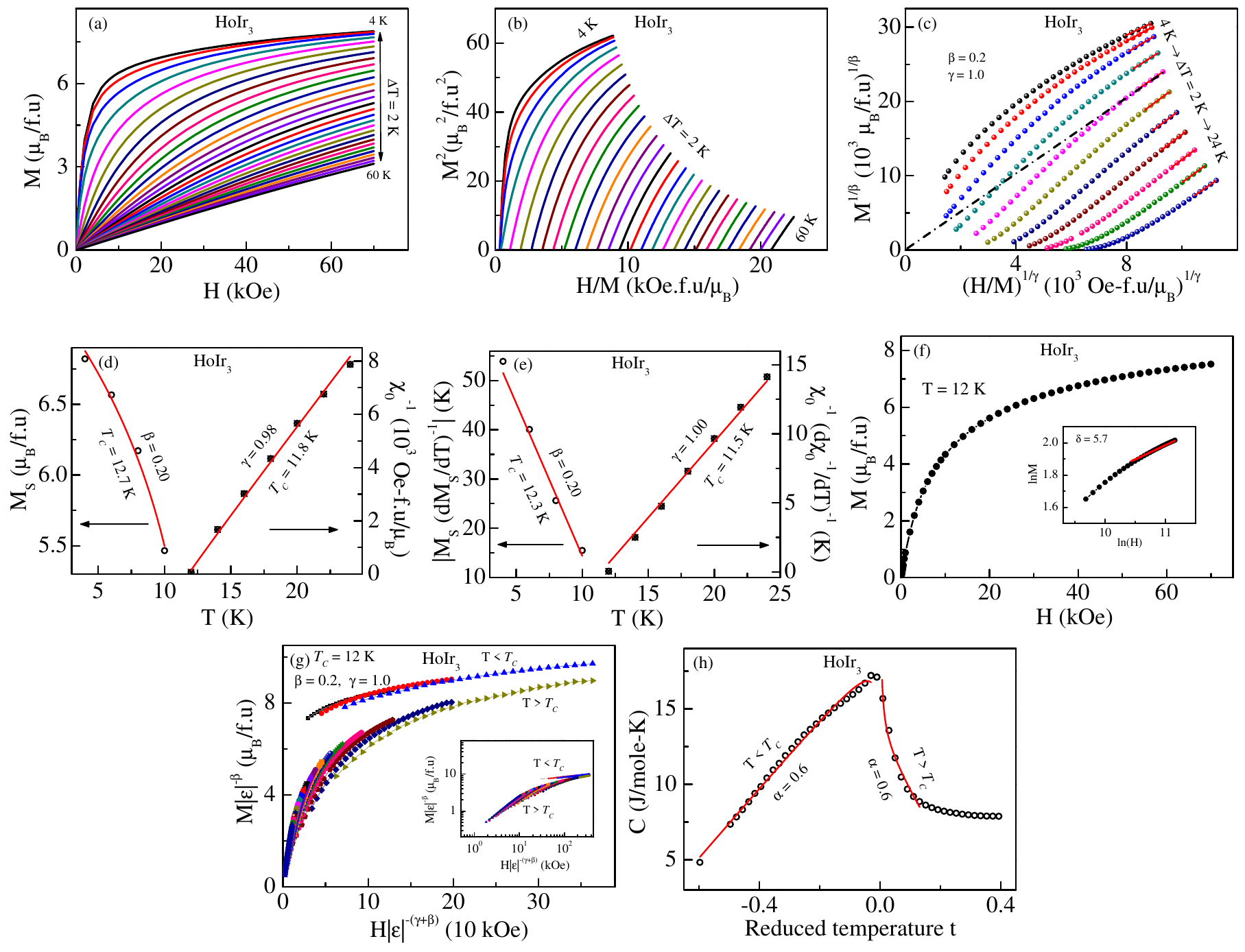}
\caption{Critical behavior of HoIr$_{3}$: (a) Isothermal magnetization curves at temperatures around ($T_{\rm{C}}$); (b) Arrott plot at different temperatures close to the Curie temperature ($T_{\rm{C}}$); (c) Modified Arrott plot. Solid lines are the linear fit of the isotherms at high temperature region. The isotherm close to the Curie temperature ($T_{\rm{C}} \sim$ 12 K) almost passes through the origin; (d) Temperature dependence of spontaneous magnetization and inverse initial susceptibility. The solid lines are the fit to the power law eqs. (\ref{eqn:powerlaw1}), (\ref{eqn:powerlaw2}); (e) Kouvel-Fisher plot of spontaneous magnetization and inverse initial susceptibility. Solid lines are the linear fit to the data; (f) Critical isotherm close to the Curie temperature ($T_{\rm{C}}$). The inset shows the same on log-log scale. The solid line is the linear fit following eq. (\ref{eqn:powerlaw3}); (g) Scaled magnetization below and above  $T_{\rm{C}}$. This plot shows that all the data collapse onto two different curves: one below $T_{\rm{C}}$ and another above  $T_{\rm{C}}$. Inset shows the same on a log scale; (h) Heat capacity data on a reduced temperature scale below and above  $T_{\rm{C}}$. The solid line is the linear fit following eq. (\ref{eqn:Critical_Spheat}).}
\label{fgr:Critical_HoIr3}
\end{center}
\end{figure*}

\subsection{Critical behavior of the magnetization and susceptibility}

In the earlier analysis, we have seen that while GdIr$_3$ exhibit a long range magnetic order, an addition glassy feature is observed in both TbIr$_3$ and HoIr$_3$. It thus raises a question whether the magnetic ordering observed in these three compounds are indeed have a long range character and if so one would be interested to know more about the universality class of these magnetic systems. These information could be extracted by studying the critical behavior of different physical properties, viz. \textit{M(H,T)}, \textit{C(T)}, \textit{etc}, around their respective magnetic transition temperatures. The critical analysis of these physical properties helps us to understand and classify a system according to the nature and strength of their respective magnetic interactions. Critical analysis study utilizes the fact that any phenomena that takes place in the vicinity of the phase transition temperature can be associated with a power law behavior of the reduced temperature ($\varepsilon$ = $\frac{T-T_{\rm C}}{T_{\rm C}}$). For example, magnetic correlation length $\xi$ can be expressed as $\xi$ = $\xi_0|\varepsilon|^{-\nu}$, where $\nu$ is known as critical exponent.\\

Following the same argument, one can express several other physical quantities viz., $M_{S}(T)$, $\chi_0(T)$, $M(H, T = T_{\rm{C}})$, \textit{C(T)}, \textit{etc.} with similar power law expression as \cite{29, 40, 32}:
\begin{eqnarray}
M_S (0, T) = M_0(-\varepsilon)^\beta \hspace{2cm} \varepsilon<0, \label{eqn:powerlaw1}\\
\chi_{0}^{-1}(0, T) = \left(\frac{h_0}{M_0}\right)(\varepsilon)^\gamma \hspace{2cm} \varepsilon<0, \label{eqn:powerlaw2}\\
M(H,T_{\rm C}) = A_0(H)^\frac{1}{\delta} \hspace{2cm} \varepsilon=0, \label{eqn:powerlaw3}\\
C(T) = C_0 \varepsilon ^{-\alpha}\hspace{3.5cm}\label{eqn:powerlaw4}
\end{eqnarray}

\noindent where $M_0$, $\frac{h_0}{M_0}$ and $A_0$, $C_0$ are the critical amplitudes, $M_{S}$ is the spontaneous magnetization, $\chi_0$ is the initial susceptibility. Depending on the characteristic of various universality classes, viz. 2D Ising model, 3D Ising model, mean field, 3D Heisenberg model, tricritical mean field, XY model \textit{etc} the critical exponents $\alpha$, $\beta$, $\gamma$ and $\delta$ can assume different set of values (see table \ref{table:6}). Conversely by carrying out critical analysis and obtaining the values of $\alpha$, $\beta$, $\gamma$ and $\delta$, one may associate the compound with the universality class it belongs to. The values of critical exponents associated with different universal class is given in table \ref{table:6}.

\begin{table}[h!]
\caption{Value of critical exponents according to different ideal model. \cite{42, 43, 43a, 50}}
\label{table:6}
\centering
\vspace{0.3cm}
\begin{tabular}{ccccc}
\hline
\hline
  & $\alpha$  &  $\beta$    & $\gamma$ & $\delta$  \\
\hline
Mean field & 0 & 0.5 & 1.0 & 3.0 \\
2D Ising & 0 & 0.12 & 1.75 & 15 \\
3D Ising & 0.11 & 0.32 & 1.24 & 4.82 \\
3D Heisenberg & -0.11 & 0.36 & 1.38 & 4.90 \\
3D XY & -0.007 & 0.34 & 1.34 & 4.8 \\
Tricritical mean field & 0.5 & 0.25 & 1.0 & 5.0 \\
\hline \hline
\end{tabular}
\end{table}

It must be pointed out here that although, the power law behavior expressed in eqs. (\ref {eqn:powerlaw1}), (\ref {eqn:powerlaw2}), (\ref {eqn:powerlaw3}), (\ref {eqn:powerlaw4}) are independent to each other, but the critical exponents are not so. The critical exponents can be linked using different scaling relations. For example, magnetization $M(H, T)$ can be expressed using two independent functions of \textit{H} and \textit{T} as,
\begin{eqnarray}
M(H,T) = F(T)\times G(T,H) \label{eqn:Scaling1}
\end{eqnarray}
\noindent where \textit{F(T)} is a function of \textit{T} alone, while \textit{G(T,H)} is a function of both \textit{T} and \textit{H}. Solving analytically one can rewrite equation \ref {eqn:Scaling1} as,
\begin{eqnarray}
M(H,\varepsilon) = (\varepsilon)^\beta f{_\pm}[H/\varepsilon^{\gamma+\beta}] \label{eqn:Scaling2}
\end{eqnarray}

\noindent where $f_+$ and $f_-$ are the functions of temperatures above and below $T_{\rm{C}}$, respectively \cite{29, 40}. Using different boundary conditions, one can obtain a scaling relationship,
\begin{equation}
\delta = 1+\left(\frac{\gamma}{\beta}\right)
\label{eqn:Widom}
\end{equation}
which is widely known as Widom scaling relation \cite{32}.\\

If the scaled or renormalized magnetization and magnetic field are defined as, $ m = |\varepsilon|^{-\beta} M(H,\varepsilon)$ and $h = |\varepsilon|^{-(\gamma+\beta)}H$, eq. (\ref{eqn:Scaling2}) reduces to a simple form\\
\begin{equation}
m = f_\pm(h)\label{eqn:Scaling4}
\end{equation}

 This equation is quite significant as it shows that with appropriate choice of a particular set of $\beta$, $\gamma$ and $\delta$ the scaled magnetization (\textit{m}) as a function of scaled field (\textit{h}) taken at different temperatures can essentially be converged to two different universal curves: $f_+(h)$ for temperatures above $T_{\rm{C}}$ and $f_-(h)$ for temperatures below $T_{\rm{C}}$. \\

 As shown in eqs. (\ref {eqn:powerlaw1}), (\ref {eqn:powerlaw2}), (\ref {eqn:powerlaw3}), (\ref {eqn:powerlaw4}) different measurements can be employed to estimate different critical exponents. For example, by studying the isothermal magnetization close to critical temperature, eq. (\ref {eqn:powerlaw3}) indicates that one can obtain information on $\delta$ (and subsequently on $\alpha$ and $\beta$). From table \ref{table:6} we see that for mean fielf like variation $\delta$ is close to 3, that is eq. (\ref {eqn:powerlaw3}) reduces to
\begin{equation}
M = A_{0}H^{1/3}
\end{equation}

\noindent where \textit{$A_{0}$} is a constant. This is generally known as Arrott equation \cite{30}. Using this equation a set of magnetic isotherms obtained experimentally near $T_{\rm C}$ can be turned into another set of parallel straight lines in the $M^2$ vs $H/M$ representation. This reconstructed magnetic isotherms are called Arrott plot \cite{30}. The magnetic isotherm of Arrott plot that passes through origin defines the $T_{\rm C}$. However, the material that does not obey mean field approximation cannot produce such set of parallel straight lines. A more generalized equation has been provided by Arrott and Noaks as \cite{34},

\begin{equation}
(H/M)^{1/\gamma}= a\left(\frac{T-T_{\rm C}}{T_{\rm C}}\right)+bM^{1/\beta}, \label{eqn:ArrottNoak}
\end{equation}
(where \textit{a} and \textit{b} are constants) which is used to obtain a set of parallel straight lines in the $M^{1/\beta}$ vs $\left(H/M\right)^{1/\gamma}$ representation. This plot obeying Arrott-Noak equation of state is often referred as modified Arrott plot \cite{34}. Thus a self consistent values of $\beta$, $\gamma$ and $\delta$ can be obtained by same set of data (isothermal magnetization) using different analytical approach as presented in eqs. (\ref{eqn:powerlaw1}), (\ref{eqn:powerlaw2}), (\ref{eqn:powerlaw3}), (\ref{eqn:Scaling2}) and (\ref{eqn:ArrottNoak}).\\

Magnetic isotherms of GdIr$_{3}$ in the temperature range 60-130 K in an interval of 2 K near $T_{\rm{C}}$ are shown in fig. \ref {fgr:Critical_GdIr3}(a). To test the applicability of Arrott equation in this system, magnetic isotherms are plotted in $M^2$ vs $H/M$ depiction in fig. \ref {fgr:Critical_GdIr3}(b). The nonlinear nature of the same plot suggests that the material does not belong to the ideal mean field family. A set of parallel straight lines however could be obtained in the temperature range 70-100 K by redrawing the Arrott plot using  eq. (\ref{eqn:ArrottNoak}) by considering $\beta$ = 0.45 and $\gamma$ = 1.1 (fig. \ref{fgr:Critical_GdIr3}(c)). We are able to obtain such set of parallel straight lines in the modified Arrott-plot over quite a large region -0.19 $<$ $\varepsilon$ $<$ 0.14 around the critical temperature. Incidentally, not many system exist where the critical region spans over such a wide temperature zone \cite{43a}. The modified Arrott plot helps us to estimate the magnetic ordering temperature far more accurately ($T_{\rm{C}} \sim$ 87 K). To test the validity of $\beta$ and $\gamma$,  we have estimated the same set of parameters using different methods as described in eqs. (\ref{eqn:powerlaw1}) \& (\ref{eqn:powerlaw2}) near the critical temperature region. Fig. \ref {fgr:Critical_GdIr3}(d) shows the extracted values of $M_S(T)$ and $\chi_{0}^{-1}(T)$ and fitted using eqs. (\ref{eqn:powerlaw1}) \& (\ref{eqn:powerlaw2}). The nonlinear fitting of $M_S$ vs. \textit{T} and $\chi_{0}^{-1}$ vs. \textit{T} suggest $\beta = 0.44$, $T_{\rm{C}} = 87.31$ K and $\gamma = 1.08$, $T_{\rm{C}} = 87.1$ K, respectively. A linear dependency with temperature can be obtained using a method suggested by Kouvel and Fisher \cite{33} where $M_S\left(dM_S/dT\right)^{-1}$ and $\chi_{0}^{-1}\left(d\chi_{0}^{-1}/dT\right)^{-1}$ are plotted (Fig. \ref {fgr:Critical_GdIr3}(e)) against temperature having slopes $1/\beta$ and $1/\gamma$ respectively. Fig. \ref {fgr:Critical_GdIr3}(e) suggests a value of $\beta = 0.45$ and $\gamma = 1.06$ for GdIr$_3$ using the Kouvel-Fisher technique.\\

The above mentioned set of analysis helps us to determine the Curie temperature of GdIr$_3$ with resonable confidence level. To estimate the other parameter $\delta$, we have chosen the magnetic isotherm measured experimentally at a temperature close to $T_{\rm{C}}$. Fig. \ref {fgr:Critical_GdIr3}(f) shows the magnetic isotherm at $T \simeq T_{\rm{C}}$ for GdIr$_3$. Inset of this figure shows logarithmic behavior of same isotherm. A linear fit of the inset data using eq. (\ref{eqn:powerlaw3}) suggests the value of $\delta = 3.33$. Widom relation \cite{32}, presented earlier in eq. (\ref{eqn:Widom}) also suggest an alternative method to estimate the value of $\delta$ when the exponents $\beta$ and $\gamma$ are known. The value of $\delta$ thus around to be 3.35, (taking the values of $\beta$ and $\gamma$ from Kouvel-Fisher technique as shown in fig. \ref {fgr:Critical_GdIr3}(e)), which is very close to that obtained earlier from different methods. \\

The values of $\beta$ and $\gamma$ can also be independently estimated by using two sets of scaling relations (eq. \ref{eqn:Scaling4}) to different magnetic isotherms, above and below $T_{\rm{C}}$, respectively. Tuning the values of $\beta$ and $\gamma$ we have been successfully able to merge all the rescaled relations into two different universal curves (Fig. \ref {fgr:Critical_GdIr3}(g)) for $T<T_{\rm{C}}$ and $T>T_{\rm{C}}$. The rescaled curves in logarithmic scale converges near $T = T_{\rm{C}}$ as shown in the inset of fig. \ref{fgr:Critical_GdIr3}(g). The parameters thus obtained also found to match with the same set of parameters estimated earlier using different methods (table \ref{table:4}).\\

Apart from magnetization, specific heat at constant pressure and in absence of external magnetic field also follow a power law at temperatures close to $T_{\rm{C}}$ (eq. (\ref{eqn:powerlaw4})). We have fitted the measured heat capacity in the critical region as a function of reduced temperature using the following eq., \cite{42}\\
\begin{equation}
\label{eqn:Critical_Spheat}
C_P^\pm = \left(\frac{A_\pm}{\alpha}\right)|\varepsilon|^{-\alpha}+B+C\varepsilon
\end{equation}
\noindent where $\alpha$ is the critical exponent, while $A_{\pm}$, \textit{B} and \textit{C} are constants. The subscript `+' is for $\varepsilon > 0$ i.e. for $T > T_{\rm{C}}$ and `-' stands for $\varepsilon < 0$ i.e. for $T < T_{\rm{C}}$. Fig. \ref {fgr:Critical_GdIr3}(h) shows the $C_P(T)$ behavior for GdIr$_3$, near its transition temperature. This data is fitted with eq. (\ref{eqn:Critical_Spheat}) below and above $T_{\rm{C}}$ and the critical exponent $\alpha$ are obtained for both the scaled curves (Table \ref{table:4}). \\

The same set of analysis described above in this section have also been carried out for other two compounds TbIr$_3$ and HoIr$_3$, and the critical exponents are obtained by all these methods (Table \ref{table:4}). The same type of figures for TbIr$_3$ and HoIr$_3$ compounds are shown in figs. from \ref {fgr:Critical_TbIr3}(a) to \ref{fgr:Critical_TbIr3}(h) and from \ref {fgr:Critical_HoIr3}(a) to \ref{fgr:Critical_HoIr3}(h) respectively. The critical exponents estimated for GdIr$_3$, TbIr$_3$ and HoIr$_3$ by different methods closely match with those reported in mean field theory, 3-D Heisenberg magnetic class and tricritical mean field theory, respectively \cite{36, 37, 38, 39, 41, 43}. Thus the above analysis suggest that while GdIr$_3$ obeys mean field theory, TbIr$_3$ belongs to 3-D Heisenberg class and HoIr$_3$ follows tricritical mean field theory. \\

 Studying the universality class of the magnetic phase transition also helps us in understanding the range of exchange interaction \textit{J(r)} \cite{35}. The renormalization group theory analysis for such systems by Fisher \textit{et al}. \cite{33} suggests that the exchange interaction, \textit{J(r)} varies as, 1/\textit{r}$^{d+\sigma}$, \cite{35} where \textit{d} is the dimension of the system and $\sigma$ is the range of the exchange interaction. For a 3D system the exchange interaction is \textit{J(r)} = 1/\textit{r}$^{3+\sigma}$ with $\dfrac{3}{2}$ $\leq$ $\sigma$ $\leq$ 2. For 3-D Heisenberg system $\sigma$ is equal to 2, thus \textit{J(r)} varies with \textit{r} as \textit{r}$^{-5}$ and the interaction strength decays fastest among all classes. The mean field exponents hold if \textit{J(r)} varies with \textit{r} as \textit{r}$^{-4.5}$ (for $\sigma$  = 3/2). In the intermediate range, the exponents belong to a different universality class which depends upon the value of $\sigma$. Thus for GdIr$_3$ sample, the interaction strength is of long range, but in case of TbIr$_3$, the interaction is of short range 3D Heisenberg-type. For intermediate values of $\sigma$, the critical exponents follow different kind of universality class such as triclinic mean field class for HoIr$_3$.

\begin{table*}[t]
\caption{Critical exponents obtained from magnetization, heat capacity and magnetocaloric data using different technique for \textit{R}Ir$_3$ (\textit{R} = Gd, Tb, Ho) compounds}
\label{table:4}
\centering
\begin{tabular}{lp{1.5cm}p{1.8cm}p{2.2cm}p{1.5cm}p{1.5cm}p{1.7cm}p{1.5cm}p{2.5cm}}
\toprule

\rotatebox{90}{Compound}               & \rotatebox{90}{Critical exponents} \rotatebox{90}{obtained from} \rotatebox{90}{modified Arrot plot }            & \rotatebox{90}{Critical exponents}\rotatebox{90}{obtained from} \rotatebox{90}{M$_S$(T) and $\chi_0^{-1}$(T)} \rotatebox{90}{curves}
            & \rotatebox{90}{Critical exponents} \rotatebox{90}{obtained from} \rotatebox{90}{Kouvel Fisher} \rotatebox{90}{method}       &       \rotatebox{90}{Critical exponents} \rotatebox{90}{obtained from} \rotatebox{90}{magnetic isotherms }        &     \rotatebox{90}{Critical exponents} \rotatebox{90}{obtained from} \rotatebox{90}{Scaling}              &      \rotatebox{90}{Critical exponent}\rotatebox{90}{obtained from} \rotatebox{90}{C$_P$(T) curve}              &    \rotatebox{90}{Critical exponents}\rotatebox{90}{obtained from} \rotatebox{90}{MCE data}    &    \rotatebox{90}{Types of} \rotatebox{90}{interaction}                                    \\ \hline
\multirow{5}{*}{GdIr$_{3}$} & $\beta= 0.45$     & $\beta= 0.44$     & $\beta= 0.45$     & \multirow{5}{*}{$\delta= 3.33$} & \textbf{$\beta= 0.45$}  & $\alpha= 0.008$          & $n = 0.64$     & \multirow{5}{*}{Mean Field}             \\
                       & $\gamma= 1.1$    & $T_{\rm{C}}= 87.31$ & $T_{\rm{C}}= 87.34$   &                            & $\gamma= 1.1$          & $(T<T_{\rm{C}})$   & $\delta= 3.38$ &                                         \\
                       & $\delta= 3.44$    & $\gamma= 1.08$    & $\gamma= 1.06$    &                            & $\delta= 3.44$          & $\alpha= 0.001$          & $\beta= 0.45$  &                                         \\
                       & $T_{\rm{C}}= 87.0$  & $T_{\rm{C}} =87.1$ & $T_{\rm{C}}= 87.0$ &                            & $T_{\rm{C}}= 87.0$       & $(T > T_{\rm{C}})$ & $\gamma= 1.07$ &                                         \\
                       &              & $\delta= 3.45$    &   $\delta= 3.35$           &                            &                    &                    &           &                                         \\ \hline
\multirow{5}{*}{TbIr$_{3}$} & $\beta= 0.32$     & $\beta= 0.32$     & $\beta= 0.32$     & \multirow{5}{*}{$\delta= 5.10$} & $\beta= 0.32$           & $\alpha= -0.114$          & $n = 0.60$     & \multirow{5}{*}{3d Heisenberg}          \\
                       & $\gamma= 1.32$    & $T_{\rm{C}} = 39.70$ & $T_{\rm{C}} = 39.83$    &                            & $\gamma= 1.33$   & $(T < T_{\rm{C}})$          & $\delta= 5.10$ &                                         \\
                       & $\delta= 5.12$    & $\gamma= 1.32$    & $\gamma= 1.32$     &                            & $\beta= 5.15$          & $\alpha= -0.12$           & $\beta= 0.32$  &                                         \\
                       & $T_{\rm{C}} = 39.0$ & $T_{\rm{C}}= 39.10$ & $T_{\rm{C}}=39.82$ &                      & $T_{\rm{C}}= 40$ & $(T > T_{\rm{C}})$          & $\gamma= 1.34$ &                                         \\
                       &     &   $\delta= 5.12$           &     $\delta= 5.12$         &                            &                    &                    &           &                                         \\  \hline
\multirow{5}{*}{HoIr$_{3}$} & $\beta= 0.20$     & $\beta= 0.20$     & $\beta= 0.20$  & \multirow{5}{*}{$\delta= 5.70$} & $\beta= 0.20$           & $\alpha= 0.6$          & $n = 0.40$     & \multirow{5}{*}{Tricritical Mean Feild} \\
                       & $\gamma= 1.0$    & $T_{\rm{C}}= 12.70$ & $T_{\rm{C}}= 12.30$    &                            & $\gamma= 1.0$   & $(T<T_{\rm{C}})$          & $\delta= 5.98$ &                                         \\
                       & $\delta= 6.0$    & $\gamma= 0.98$    & $\gamma= 1.0$    &                            & $\delta= 6.0$         & $\alpha= 0.6$           & $\beta= 0.21$  &                                         \\
                       & $T_{\rm{C}}= 12.0$ & $T_{\rm{C}}= 11.8$ & $T_{\rm{C}}= 11.5$ &                            & $T_{\rm{C}}= 12.0$ & $(T > T_{\rm{C}})$          & $\gamma= 1.04$ &                                         \\
                       &     &    $\delta= 5.9$          &   $\delta= 6.0$           &                            &                    &                    &           &\\

\hline
\end{tabular}
\end{table*}


\begin{figure*}
\begin{center}
\includegraphics[scale=1.0]{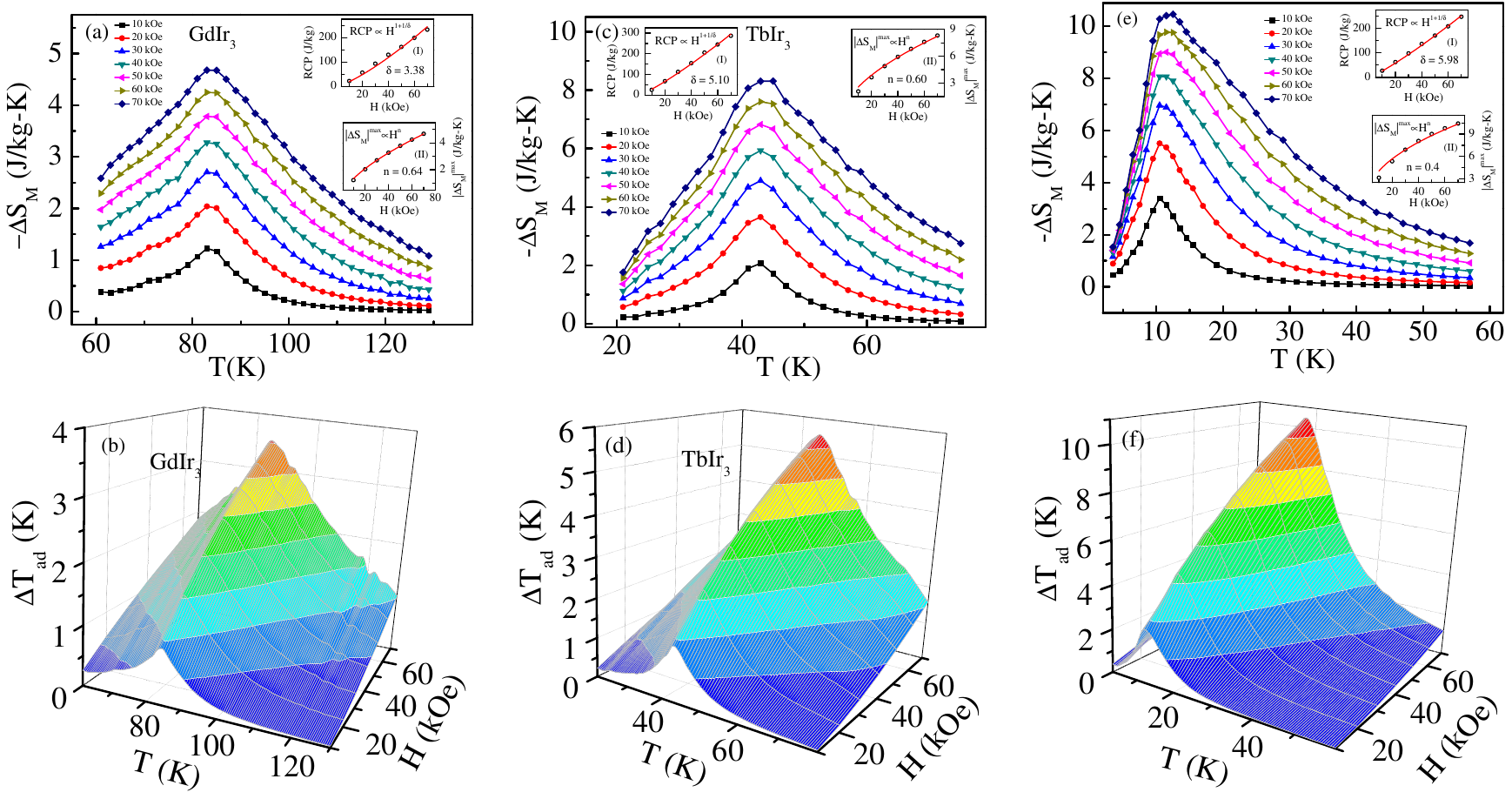}
\caption{Temperature dependence of isothermal magnetic entropy change and temperature dependence of the adiabatic temperature change of GdIr$_3$ ((a), (b)), TbIr$_3$ ((c), (d)) and HoIr$_3$ ((e), (f)) compounds respectively. The insets (I) of (a), (c) and (e) show the variation of \textit{RCP} as a function of \textit{H} and insets (II) represents the variation of $|\Delta S_{M}|^{max}$ as a function of \textit{H} of compounds GdIr$_3$, TbIr$_3$ and HoIr$_3$. The solid lines show the power law fits of equations \ref{eqn:n} and \ref{eqn:RCP} of the inset curves.}
\label{fgr:MCE}
\end{center}
\end{figure*}

\begin{figure}
\begin{center}
\includegraphics[scale=0.24]{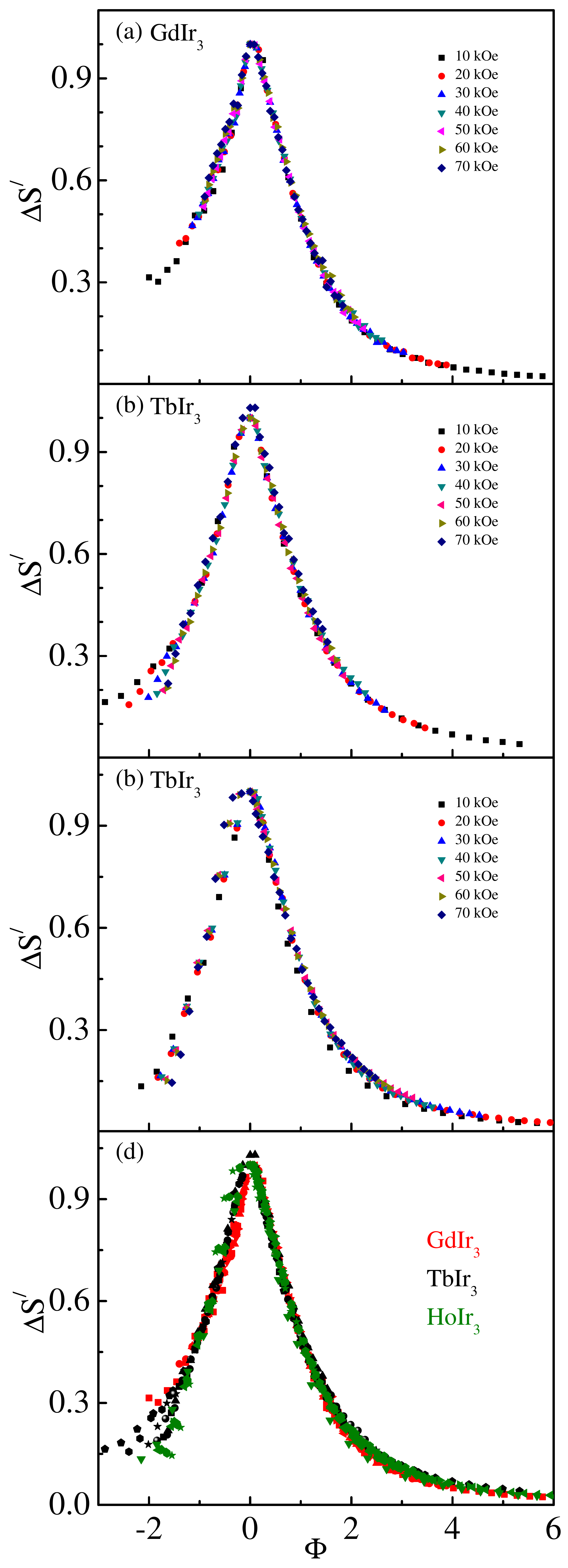}
\caption{Universal curves for (a) GdIr$_3$, (b) TbIr$_3$, (c) HoIr$_3$ compounds respectively obtained using two temperatures of reference and (d) universal curve for three compounds altogether.}
\label{fgr:Universal}
\end{center}
\end{figure}

\subsection{Magnetocaloric Effect}

In the preceeding section, we have discussed about the nature of magnetic interaction present in these three $\textit{R}$Ir$_3$ ($\textit{R}$ = Gd, Tb, Ho) compounds through critical analysis study. It was concluded that GdIr$_3$ follows  the mean field theory, TbIr$_3$ belongs to the 3D Heisenberg magnetic class while HoIr$_3$ obeys tricritical mean field theory. The validity of such conclusion can also be checked through the study of magnetocaloric effect (MCE).\\

MCE is an environment friendly and alternative technique over the conventional gas compression/expansion methods to achive cooling process. It is defined as a change in the temperature (heating or cooling) of materials due to the application of a magnetic field and can be estimated either through heat capacity or isothermal magnetization measurements.\\

In the later procedure, the magnetic entropy change, $\Delta S_M$ using Maxwell relation \cite{46} is defined as

\begin{equation}
\Delta S_M(T,H)= \int_0^H\left[\frac{\partial M}{\partial T}\right]dH \label{eqn:DeltaS_M}
\end{equation}

\noindent $\Delta$S$_M (T, H)$ have been obtained from a set of isothermal magnetization data by solving eq. (\ref{eqn:DeltaS_M}) using numerical approximation method. Figs. \ref{fgr:MCE} (a), (c), (e) show the magnetic entropy changes as a function of temperature for different field changes of \textit{R}Ir$_3$ (\textit{R} = Gd, Tb, Ho) compounds, respectively. The maximum values of $|$$\Delta S_M$$|$ are 4.7 J/kg-K, 8.3 J/kg-K, 10.5 J/kg-K for GdIr$_3$, TbIr$_3$, HoIr$_3$ at temperatures 83 K, 43 K, 11 K, respectively for a field change of 0 $\rightarrow$ 70 kOe respectively [fig. \ref{fgr:MCE} (a), (c), (e)]. Similar values of $\Delta S_M$ are also reported for different intermetallic compounds in this temperature range subject to similar field change \cite{57, 58}.\\

Along with magnetic entropy change, the amount of heat transfer between the hot and cold reserviors in an ideal refrigerant cycle of the material is quantified by relative cooling power (\textit{RCP}). \textit{RCP} for a particular field \textit{H} is defined as the product of maximum entropy change ($\Delta S_M$) and full width at half-maximum ($\delta T_{FWHM}$) of $|\Delta S_M|$-\textit{T} curve \cite{47}. A large value of \textit{RCP} can be achieved either by getting large $\Delta S_M$ or widespread of $\Delta S_M$ over a large temperature range, or both. \textit{RCP} values for compounds GdIr$_3$, TbIr$_3$, HoIr$_3$ have been estimated to be 232 J/kg, 287 J/kg and 248 J/kg respectively at field of 70 kOe indicating their appropriate usage in cooling technology.\\

However, to judge the applicability of a good MCE material, another important parameter is adiabatic temperature change ($\Delta T_{ad}$), that is defined as

\begin{equation}
\Delta T_{ad}=[T(S,H)-T(S,0)]_S,
\end{equation}

\noindent where \textit{T(S,H)} and \textit{T(S,0)} are the temperatures at applied field \textit{H} and no applied field respectively, for a particular entropy \textit{S}. Using thermodynamic relation, it can be written as

\begin{equation}
\Delta T_{ad}(T,H)=-\int_0^H\left[\frac{T}{C}\right]\left[\frac{\partial M}{\partial T}\right]dH
\end{equation}

\noindent where \textit{C} is the specific heat of the system at zero applied field. We have estimated $\Delta T_{ad}$ using $\Delta S_M$ and the zero field heat capacity data. The maximum values of $\Delta T_{ad}$ are 3.5 K, 5.6 K, 10.5 K for GdIr$_3$, TbIr$_3$, HoIr$_3$ at temperatures 83 K, 43 K, 11 K, respectively, at a field change of 0 $\rightarrow$ 70 kOe [fig. \ref{fgr:MCE} (b), (d), (f)]. One may notice an appreciably large value of $\Delta T_{ad}$ in HoIr$_3$ in this temperature region and similar field sweep \cite{25}. The values of $|\Delta S_{M}|^{max}$, \textit{RCP}, $\Delta T_{ad}$ for fielf change 0 $\rightarrow$ 70 kOe are represented in table \ref{table:5} for \textit{R}Ir$_3$ (\textit{R} = Gd, Tb, Ho) compounds. \\

\begin{table}[ht]
\caption{$|\Delta S_{M}|^{max}$, $\Delta T_{ad}$ (for a fielf change of 0 $\rightarrow$ 70 kOe) and \textit{RCP} for compounds \textit{R}Ir$_3$ (\textit{R} = Gd, Tb, Ho)}
\label{table:5}
\centering
\vspace{0.3cm}
\begin{tabular}{llll}
\hline
\hline
Compound  &  $|\Delta S_{M}|^{max}$   &  \textit{RCP}    & $\Delta T_{ad}$  \\
 &  J/kg-K  &  J/kg  &  K  \\
\hline
GdIr$_3$ & 4.7 & 232 & 3.5 \\
 TbIr$_3$ & 8.3 & 287 & 5.6 \\
 HoIr$_3$ & 10.5 & 248 & 10.5 \\
\hline \hline
\end{tabular}
\end{table}

As mentioned earlier in this section, MCE can also be used independently to determine the critical exponents by studying the variation of $\Delta S_M$ and \textit{RCP} as a function of applied magnetic field \cite{59}.\\

It is generally found, the field dependence of the magnetic entropy change at the critical temperature associated with a second order magnetic phase transition follows the relation \cite{49}

\begin{equation}
|\Delta S_M|^{max}\propto H^n \label{eqn:n}
\end{equation}

\noindent and the \textit{RCP} varies as \cite{50}

\begin{equation}
\rm RCP \propto H^{1+1/\delta}\label{eqn:RCP}
\end{equation}

The insets in fig. \ref{fgr:MCE} (a), (c), (e) shows the plot of $|\Delta S|^{max}$ as a function of \textit{H} and plot of\textit{ RCP} as a funtion of \textit{H}. The power law fit of eqs. (\ref{eqn:n}) and (\ref{eqn:RCP}) give the values of \textit{n} and $\delta$ for the three compounds which are shown in table \ref{table:4}.

The parameter \textit{n} in eq. (\ref{eqn:n}) has been expressed in terms of $\beta$ and $\gamma$ as, \cite{51}

\begin{equation}
n = 1+\frac{(\beta-1)}{(\beta+\gamma)}\label{eqn:Franco}
\end{equation}

Using both eqs. (\ref {eqn:Franco}) and (\ref{eqn:Widom}), $\beta$ and $\gamma$ can be expressed as a function of \textit{n} and $\delta$ as \cite{59}

\begin{eqnarray}
\beta = \frac{1}{\delta(1-n)+1} \label{eqn:Beta}\\
\gamma = \frac{(\delta-1)}{\delta(1-n)+1} \label{eqn:Gamma}
\end{eqnarray}

Using eqs. (\ref{eqn:Beta}), (\ref{eqn:Gamma}) and (\ref{eqn:Widom}) the values of $\beta$, $\gamma$ and $\delta$ are estimated for \textit{R}Ir$_3$ ($\textit{R}$ = Gd, Tb, Ho) compounds. The obtained values are displayed in table \ref{table:4} and closely matches with the values acquired by other methods discussed in section G. The closeness of the critical exponents ($\beta$, $\gamma$, $\delta$) suggest the self consistency of the analysis.\\

Another interesting feature of MCE is that if the magnetic entropy change, $\Delta S_M$ can be described in terms of appropriately chosen reduced temperature scale for any particular magnetic field, the shape and value of $\Delta S_M(T,H)$ curves for any arbitrary magnetic field can be generated, even without knowing the critical exponents, subject to an assumption that no major change in magnetic interaction takes place abruptly in the system \cite{51,53, 54, 55,56}. Additionally, the same universal curve can also be used to predict the $\Delta S_M(T,H)$ curves for other members of same series of compounds, that may be having similar nature of magnetic interaction and magnetic ordering temperature is known \cite{53}. Thus, the utility of such universal or master curve is to use to extrapolate data in the temperature ranges where the sample was not even measured \cite{53, 56}.\\

In general, such a master curve is obtained by normalizing the  parameter $\Delta S^{\prime} =  |\Delta S_{M}|/|\Delta S_{M}|^{max}$, as a function of rescaled temperature \cite{47, 51}

\begin{equation}
\Phi = \frac{(T-T_{\rm C})}{(T_r-T_{\rm C})}
\end{equation}

\noindent where, $T_r$ is the temperature at which $|\Delta S_{M}| = a \times |\Delta S_{M}|^{max}$ (`\textit{a}' is an adjustable parameter which can take value between 0 and 1). However, if the sample is magnetically inhomogeneous or the measuring field is too low, then one need to use two scaling parameters $T_{r1}$ and $T_{r2}$ instead of a single one ($T_r$) \cite{47, 51}. Therefore temperature, $\Phi$ is defined as \cite{43, 47, 51}

\begin{eqnarray}
\Phi =
\begin{cases}
 -\frac{(T-T_{\rm{C}})} {(T_{r1}-T_{\rm{C}})}, &       T \leq T_{\rm{C}}\\
        \hspace{.3cm}\frac{(T-T_{\rm{C}})} {(T_{r2}-T_{\rm{C}})}, &T > T_{\rm{C}} \label{eqn:Universal}
        \end{cases}
\end{eqnarray}

For TbIr$_3$, GdIr$_3$ and HoIr$_3$ compounds, after rescaling the temperature using eq. (\ref{eqn:Universal}) and choosing \textit{a} = $\frac{1}{2}$, we find that $\Delta S^{\prime}(\Phi$) for all the applied fields collapse in single curves for each individual samples [fig. \ref{fgr:Universal} (a), (b), (c)].\\

Furthermore, if these three universal curves corresponding to the three different compounds, GdIr$_3$, TbIr$_3$, HoIr$_3$ are plotted together then it is found that they overlap for positive $\Phi$ $>$ 0 that is in the paramagnetic region (T $>$ \textit{T}$\rm_{C}$), while they slightly differ from each other for $\Phi$ $<$ 0 that is within the ordered region (T $<$ \textit{T}$\rm_{C}$) [fig. \ref{fgr:Universal}]. This property allows the prediction of $|$$\Delta$S$_M$$|$(T) curves even in other compounds having related compositions \cite{53}.\\

\subsection{Summary}

In summary, we report the successful synthesis of three compounds GdIr$_3$, TbIr$_3$, HoIr$_3$ which found to form in two polymorphic phases (C15b, AuCu$_3$). The dc magnetization measurements show that these compounds orders ferromagnetically, while magnetic entropy calculation from heat capacity measurement indicates that C15b phase is responsible for ferromagnetic ordering and AuCu$_3$ phase remain paramagnetic down to 2 K. The ac susceptibility measurement and time dependent relaxation measurement indicates the presence of glassy nature in TbIr$_3$ and HoIr$_3$ but is absent in GdIr$_3$. The detailed study of dynamical scaling of ac susceptibility, magnetic relaxation and memory effect measurements established both TbIr$_3$ and HoIr$_3$ to be canonical spin glass in nature. The modified Arrott plot, M$_S$(T), $\chi_{0}^{-1}(T)$ curves, Kouvel-Fisher method and specific heat analysis confirm that GdIr$_3$ obeys mean field theory, TbIr$_3$ lies in 3-D Heisenberg universality class and HoIr$_3$ follows tricritical mean field theory. The aforesaid critical analysis complies with the MCE studies. The $\Delta$T$_{ad}$ value of HoIr$_3$ found to be quite appreciable.

\subsection*{Acknowledgements}

RNB thanks to CIF, Pondicherry University for ac susceptibility measurements. The work has been carried out by the CMPID project at SINP and funded by Department of Atomic Energy, Govt. of India. We thank Dr. Santanu Pakhira for his help during data analysis.


\end{document}